\documentclass[%
nofootinbib,
 amsmath,amssymb,
 aps,
]{revtex4-2}

\usepackage{graphicx}% Include figure files
\usepackage{dcolumn}% Align table columns on decimal point
\usepackage{bm}% bold math
\usepackage{hyperref}% add hypertext capabilities
\usepackage{aas_macros}

\usepackage{orcidlink}

\def\vect#1{\mbox{\boldmath $#1$}}
\def\pardif#1#2{\frac{\partial #1}{\partial #2}}
\def\pardel#1#2{\frac{\partial #1}{\partial #2}}
\def\cc{\text{c.c.}}
\usepackage{color}
\newcommand{\wiadd}[1]{{#1}}

\begin{document}

\preprint{APS/123-QED}

\title{Parametric decay instability of circularly polarized Alfv\'en wave in magnetically dominated plasma}% Force line breaks with \\

\author{Wataru Ishizaki\orcidlink{0000-0002-7005-7139}}%
 \email{wataru.ishizaki@yukawa.kyoto-u.ac.jp}
\affiliation{Astronomical Institute, Graduate School of Science, Tohoku University, Sendai 980-8578, Japan}
\affiliation{Frontier Research Institute for Interdisciplinary Sciences, Tohoku University, Sendai 980-8578, Japan}
\affiliation{%
 Center for Gravitational Physics, Yukawa Institute for Theoretical Physics, Kyoto University, Kitashirakawa Oiwake-cho, Sakyo, Kyoto 606-8502, Japan 
}%
 \author{Kunihito Ioka\orcidlink{0000-0002-3517-1956}}%
 \email{kunihito.ioka@yukawa.kyoto-u.ac.jp}
\affiliation{%
 Center for Gravitational Physics, Yukawa Institute for Theoretical Physics, Kyoto University, Kitashirakawa Oiwake-cho, Sakyo, Kyoto 606-8502, Japan 
}%

% \collaboration{MUSO Collaboration}%\noaffiliation

%  \homepage{http://www.Second.institution.edu/~Charlie.Author}
% \affiliation{
%  Second institution and/or address\\
%  This line break forced% with \\
% }%
% \affiliation{
%  Third institution, the second for Charlie Author
% \affiliation{%
%  Authors' institution and/or address\\
%  This line break forced with \textbackslash\textbackslash
% }%

% \collaboration{CLEO Collaboration}%\noaffiliation

\date{\today}% It is always \today, today,

\begin{abstract}
% We investigate the parametric instability of circularly polarized Alfv\'en wave in a relativistically magnetized plasma, in which the energy density of the electromagnetic field can be larger than the rest mass and internal energy densities of the plasma.
% We analyzed the parametric instability of Alfv\'en waves by using the magnetohydrodynamic equations, focusing on waves whose frequency is sufficiently lower than both the plasma and cyclotron frequency.
% Analytical formulae for the dispersion relation, the conditions for the onset of instability, and the growth rate of the instability with general magnetization are obtained for the case of small wave amplitudes.
% We confirm that parametric Alfv\'en wave instabilities occur even in plasmas with relativistic magnetization.
% We also investigate the dependence of the growth rate on the magnetization and sound velocity, and find that the growth rate of the instability is a decreasing function of the magnetization, and that the parametric instability disappears in the limit of strong magnetization.
% Alfv\'en wave propagation in relativistically magnetized plasmas is a process that is ubiquitous in the context of high-energy astrophysics and should be a clue for understanding them, for example, high-energy phenomena around compact objects.
We investigate parametric decay instability (PDI) of circularly polarized Alfv\'en wave into daughter acoustic wave and backward Alfv\'en wave in magnetically-dominated plasma, in which the magnetization parameter $\sigma$ (energy density ratio of background magnetic field to matter) exceeds unity.
We analyze relativistic magnetohydrodynamics (MHD), focusing on wave frequencies sufficiently lower than the plasma and cyclotron frequencies.
We derive analytical formulae for the dispersion relation and growth rate of the instability as a function of the magnetization $\sigma$, wave amplitude $\eta$, and plasma temperature $\theta$.
We find that PDI persists even in high magnetization $\sigma$, albeit with a decreased growth rate up to $\sigma\to\infty$.
Our formulae are useful for estimating the decay of Alfv\'en wave into acoustic wave and heat in high magnetization $\sigma$ plasma, which is a ubiquitous phenomenon such as in pulsars, magnetars, and fast radio bursts.
% \begin{description}
% Secondary publications and information retrieval purposes.
% You may use the \texttt{description} environment to structure your abstract;
% use the optional argument of the \verb+\item+ command to give the category of each item. 
% \end{description}
\end{abstract}

\maketitle

%\tableofcontents

\section{Introduction} \label{sec:level1}

Fast radio bursts (FRBs) are the most luminous radio transients in the universe with millisecond duration \cite{2007Sci...318..777L,2013Sci...341...53T,2019ARA&A..57..417C,2019A&ARv..27....4P,2022A&ARv..30....2P, 2022Sci...378.3043B}.
The identification of the host galaxies \cite{2017Natur.541...58C,2017ApJ...834L...7T,2017ApJ...834L...8M,2019Sci...365..565B,2019Natur.572..352R,2020ApJ...899L...6L,2020ApJ...903..152H,2021ApJ...917...75M,2022AJ....163...69B} and the excess of the dispersion measure over the free electron density in the Galaxy \cite{2007Sci...318..777L,2013Sci...341...53T,2019ARA&A..57..417C} have shown that almost all of the sources are at cosmological distances.
FRBs have very high brightness temperatures, suggesting that such emission is due to a coherent mechanism, i.e., a nonlinear process in which many particles emit photons of the same phase simultaneously, rather than a superposition of individual particle emissions \cite{1969ApJ...155L..71K,2007A&A...463..145T,2015MNRAS.446.3687P,2021Univ....7...56L}.
The mechanism of such high-intensity emission is still unresolved and remains an important problem in astrophysics \cite{2017RvMPP...1....5M}.

The mechanism and the source of FRBs are not known \cite{2021Univ....7...56L,2020Natur.587...45Z,2023RvMP...95c5005Z}.
The discovery of an FRB-like burst, FRB200428, coinciding with an X-ray burst from the Galactic magnetar SGR 1935+2154 has made highly magnetized neutron stars promising candidates for the source objects of FRBs \cite{2020ApJ...898L..29M,2021NatAs...5..408Y,2021NatAs...5..401T,2021NatAs...5..378L,2021NatAs...5..372R,2022ApJ...931...56L,2022ApJS..260...24C}.
Two models are intensively debated regarding the emission region.
One is that the origin of FRBs is the maser emission in strongly magnetized relativistic shock waves generated outside the magnetosphere \cite{2014MNRAS.442L...9L,2017ApJ...843L..26B,2019MNRAS.485.4091M}.
It has been pointed out that this model has the difficulty that the model conflicts with the observation of the polarization \cite{2018ApJ...863....2G,2018Natur.553..182M,2019MNRAS.488..868O} and the short time variability in some bursts \cite{2020MNRAS.498..651B,2022MNRAS.510.1867L}.
The other model is that the energy of the magnetic field of the neutron star is released and converted into emission inside the magnetosphere \cite{2016MNRAS.457..232C,2017MNRAS.468.2726K,2018ApJ...868...31Y,2018MNRAS.477.2470L,2020MNRAS.494.2385K}.
Since the energy density of magnetic fields is strong near the surface of stars, magnetic reconnection \cite{1995MNRAS.275..255T,1996ApJ...473..322T,2003MNRAS.346..540L,2013ApJ...774...92P,2019MNRAS.484L.124C,2019MNRAS.485.4091M,2020ApJ...896..142B,2022ApJ...932L..20M} or starquake \cite{1989ApJ...343..839B,1995MNRAS.275..255T,2020MNRAS.494.2385K,1996ApJ...473..322T,2002ApJ...574..332T} at the surface of stars are discussed as possible energy sources.
Even though we consider phenomena occurring within the magnetosphere of a neutron star, the emission site remains uncertain.
If FRBs are generated in the magnetosphere near the stellar surface, the difficulty has been pointed out that their highly intense photons cannot escape from the magnetosphere \cite{1975Ap&SS..36..303B,1978MNRAS.185..297W,2008ApJ...682.1443L,2018MNRAS.477.2470L,2020ApJ...897....1L,2020MNRAS.494.1217K,2021ApJ...922L...7B,2023arXiv231002306N}.
Therefore, the energy released at the stellar surface needs to be transported to an outer emission region and converted into radio waves \cite{2020MNRAS.494.2385K,2020ApJ...900L..21Y,2022ApJ...933..174Y,2022MNRAS.515.2020Q}.

On the surface of the neutron star, the energy should be released in the form of magnetic pulse (including Aflv\'en wave) \cite{2020MNRAS.494.2385K,2020ApJ...897....1L,2020ApJ...900L..21Y,2022ApJ...933..174Y,2022ApJ...934..140B,2022ApJ...932L..20M} or thermalized fireball \cite{2020ApJ...904L..15I,2021ApJ...919...89Y,2022MNRAS.511.3138Y,2023MNRAS.519.4094W}.
This paper specifically examines the cases where the release occurs in the form of Alfv\'en wave. 
Fluctuations in the magnetic field propagate through the magnetosphere as Alfv\'en waves (although fast magnetosonic waves are also conceivable, they can traverse the magnetic field and, thus, should disperse). 
In the neutron star magnetosphere, the energy density of the magnetic field significantly surpasses that of the plasma, so the Alfv\'en waves are expected to heat the plasma to relativistic energy as they dissipate, and then, to generate a bunch of charged particles, which might lead to coherent emission via some plasma instabilities \cite{2020MNRAS.494.2385K,2020ApJ...900L..21Y}.
This expectation has motivated several FRB models using Alfv\'en waves.
Since the amplitude of Alfv\'en waves in the magnetosphere increases as they propagate outward, nonlinear effects during the propagation are important \cite{2020ApJ...897....1L,2020MNRAS.494.2385K,2020ApJ...900L..21Y,2022ApJ...933..174Y,2022MNRAS.515.2020Q}. 
It remains a non-trivial question whether these large amplitude waves can traverse the magnetosphere without significant dissipation before reaching the FRB emission region. 
% Consequently, it is necessary to consider the propagation of Alfv\'en waves in a strongly magnetized relativistic plasma.

Finite amplitude Alfv\'en waves decay through the excitation of daughter acoustic waves and daughter Alfv\'en waves \cite{1969npt..book.....S,1978ApJ...219..700G,1978ApJ...224.1013D}.
% This process is called as parametric decay instability which is one of the three-wave resonant interactions, and in plasma physics, it is known as induced Brillouin scattering. 
This process is one of the parametric instability, particularly one of three-wave resonant interactions.
In the context of plasma physics, it is known as induced Brillouin scattering.
In this paper, we call this process parametric decay instability (PDI).
In the non-relativistic regime, this phenomenon has been extensively explored in the context of the solar coronal heating problem and the mechanism for launching the solar wind \cite{1978RvGSP..16..689H}.
% The mechanism involves Alfv\'en waves generated in the chromosphere heating the plasma by dissipating the daughter acoustic waves, which are generated via the decay instability while the waves propagate through the corona \cite{2005ApJ...632L..49S,2006JGRA..111.6101S,2018ApJ...860...17S}. 
When Alfv\'en waves propagate from the chromosphere into the corona, the PDI generates daughter acoustic waves, which immediately heat the coronal plasma through some dissipative process (e.g., shock heating) \cite{2005ApJ...632L..49S,2006JGRA..111.6101S,2018ApJ...860...17S}.
In the magnetosphere of a neutron star, in contrast to solar physics, a magnetic field energy density is much larger than the matter energy density \cite{1983AIPC..101..163A,2001ApJ...547..437L,2010MNRAS.406.1379M,2010MNRAS.408.2092T,2019ApJ...877...53H}. 
Although such systems are often analyzed using the force-free approximation \cite{1998PhRvD..57.3219T,2019MNRAS.483.1731L,2019ApJ...881...13L}, this approach is inadequate for discussing Alfv\'en wave stability because this formulation lacks acoustic waves (i.e., slow waves). 
Additionally, within the force-free approximation, an Alfv\'en wave alone does not decay into lower-frequency Alfv\'en or fast waves through three-wave interaction (except for the interaction with $\omega=0$ waves \cite{1995ApJ...447..706M,1996ApJ...465..845N,2021JPlPh..87f9014T,2021ApJ...908..176Y}
). 
Plasma heating due to the decay of the Alfv\'en wave is also considered important in the context of magnetar X-ray bursts (e.g., \cite{1995MNRAS.275..255T,1998PhRvD..57.3219T,2019ApJ...881...13L}, and in particular, \cite{2014ApJ...787...84T} for calculation without force-free approximation\footnote{
Their calculation \cite{2014ApJ...787...84T} is mainly focused on the macroscopic view of how much the plasma in the magnetosphere can be heated, rather than on the fundamental processes of the decay of the Alfven wave, which is our focus.
See Section \ref{sec:gensig} for a comparison of their results with ours.
}).
Consequently, an investigation of the stability of Alfv\'en waves in magnetically-dominated plasmas, without imposing the force-free approximation, is essential.

The PDI in magnetically-dominated plasmas remains largely unexplored.  
Matsukiyo \& Hada (2003) \cite{2003PhRvE..67d6406M} investigated the stability of Alfv\'en waves in two-fluid pair plasmas, providing analytical expressions for the growth rate of the instability. 
% However, their calculation is limited to the weakly relativistic regime, so that it is insufficient for application to the magnetosphere of neutron stars.
However, as we will see later, the expression they obtained is inaccurate when the sound speed of plasma is sufficiently slow or fast (see equation (\ref{eq:growthrate_gen}) and the following text for details).
Lopez et al. (2012) \cite{2012PhPl...19h2104L} incorporated full relativistic effects, but their work lacks an analytical discussion of the instability growth rate, impeding a systematic understanding. 
Deeply in the neutron star magnetosphere, the plasma frequency $\omega_p$ and cyclotron frequency $\omega_c$ significantly exceed the frequency of interest, even in comparison to the typical frequency of FRB emission $\omega_0\sim10^9~{\rm Hz}$ \cite{2018MNRAS.477.2470L,2020ApJ...904L..15I,2019MNRAS.483.4175Y},
\wiadd{
namely
%%%%%%%%%%%%%%%%%%%%%%%%%%%%%%%%%%%%%%%%%
\begin{equation}\label{eq:freq:plasma}
    \omega_p=\sqrt{\frac{4 \pi \mathcal{M} n_{\mathrm{GJ}} e^2}{m_e}} \sim 4.7 \times 10^{11} \mathrm{~Hz} ~\mathcal{M}^{1 / 2}\left(\frac{P}{1~\mathrm{s}}\right)^{-1 / 2}\left(\frac{B_{\mathrm{NS}}}{10^{15}~\mathrm{G}}\right)^{1 / 2},
\end{equation}
%%%%%%%%%%%%%%%%%%%%%%%%%%%%%%%%%%%%%%%%%
%%%%%%%%%%%%%%%%%%%%%%%%%%%%%%%%%%%%%%%%%
\begin{equation}\label{eq:freq:cyclrotron}
    \omega_c=\frac{e B_{\mathrm{NS}}}{m_e c} \sim 1.8 \times 10^{22} \mathrm{~Hz}~\left(\frac{B_{\mathrm{NS}}}{10^{15}~ \mathrm{G}}\right),
\end{equation}
%%%%%%%%%%%%%%%%%%%%%%%%%%%%%%%%%%%%%%%%%
where $c$ is the speed of light, $e$ is the unit charge of an electron, $m_e$ is the electron mass, $P$ is the rotation period of the neutron star, $B_{\rm NS}$ is the magnetic field strength on the surface of the neutron star, and $n_{\rm GJ}$ is the Goldreich-Julian density, which is the lower limit of particle number density in the neutron star magnetosphere \cite{1969ApJ...157..869G}. $\mathcal{M}\gtrsim1$ is multiplicity, the ratio of the actual number density to $n_{\rm GJ}$, the value of which is highly uncertain, depending on the details of the pair production process in the magnetosphere, but is theoretically and observationally estimated to be between $\mathcal{O}(1)$ and $10^7$ \cite{2011MNRAS.410..381B,2013MNRAS.429.2945T,2006ApJ...652.1475H,2015ApJ...810..144T,2019ApJ...871...12T,2020ApJ...904L..15I,2020ApJ...896..142B}.
Here, we estimate the characteristic frequencies using the electron mass, because the particles that consist of plasma in the neutron star magnetosphere are thought to be dominated by electrons and positrons due to the electromagnetic cascade \cite{1969ApJ...157..869G}.}
\wiadd{Since these characteristic frequencies (\ref{eq:freq:plasma}) and (\ref{eq:freq:cyclrotron}) are much higher than the frequency of interest $\omega_0\lesssim10^9~{\rm Hz}$}, the magnetohydrodynamic (MHD) model, rather than the two-fluid model, is easy to handle for describing Alfv\'en wave propagation in the neutron star magnetosphere.

In this paper, we study the stability of Alfv\'en waves in magnetically-dominated plasmas based on the MHD model, aiming to achieve a systematic understanding of the timescale of the instability in the neutron star magnetosphere.
We investigate the stability of Alfv\'en waves propagating parallel to the background magnetic field in the fluid's rest frame.
For simplicity, we analyze the 1-D case (i.e., both parent and daughter waves are propagating parallel to the background magnetic field) and focus on circularly polarized plane waves as parent Alfv\'en waves, which are easy to handle in analytical treatment due to the existence of exact analytical solutions.
We determine the growth rate of the PDI and elucidate its dependence on system parameters, particularly the magnetization parameter $\sigma$ (the ratio of magnetic to matter energy density).
The structure of this paper is as follows:
In Section 2, we formulate the governing equations for perturbations added to the exact solution of Alfv\'en waves with finite amplitude.
Section 3 derives the dispersion relation from the perturbation equations, leading to the determination of the growth rate of the PDI.
Section 4 summarizes the obtained results, discusses the interpretation and application of our findings, and outlines future prospects.
\wiadd{Unless otherwise noted, we use the Gaussian-cgs unit system in this paper.}

\section{Formulation}

\subsection{Basic equations}
In this paper, we analyze the stability of the Alfv\'en wave in the initially rest frame of plasma (we call it the reference frame).
We assume that the frequencies of waves considered in this paper are sufficiently smaller than both the cyclotron frequency and the plasma frequency, and that the plasma is quasi-neutral.
Therefore, we adopt the ideal magnetohydrodynamic (MHD) equation as the governing equation.
Furthermore, for simplicity, in this paper, we only consider the case where the waves propagate in a direction parallel to a constant magnetic field oriented in the $z$ direction.
When there is no change of physical quantities in the $x$ and $y$ directions (i.e., we consider a 1D problem), the special relativistic MHD equations can be written as follows:
%%%%%%%%%%%%%%%%%%%%%%%%%%%%%%%%%%%%%%%%%
\begin{equation}\label{eq:basic:energy}
    \frac{1}{c}\frac{\partial}{\partial t}\left[(\epsilon+p) \gamma^2-p+\frac{1}{8 \pi}\left(E^2+B^2\right)\right]+\frac{\partial}{\partial z}\left[(\epsilon+p) \gamma^2 \beta_z+\frac{1}{4 \pi}(\boldsymbol{E} \times \boldsymbol{B})_z\right]=0,
\end{equation}
%%%%%%%%%%%%%%%%%%%%%%%%%%%%%%%%%%%%%%%%%
%%%%%%%%%%%%%%%%%%%%%%%%%%%%%%%%%%%%%%%%%
\begin{equation}\label{eq:basic:momz}
 \frac{1}{c}\frac{\partial}{\partial t}\left[(\epsilon+p) \gamma^2 \beta_z+\frac{1}{4 \pi}(\boldsymbol{E} \times \boldsymbol{B})_z\right]+\frac{\partial}{\partial z}\left[(\epsilon+p) \gamma^2 \beta_z^2-\frac{1}{4 \pi}\left(E_z^2+B_z^2\right)\right]+ \frac{\partial}{\partial z}\left[p+\frac{E^2+B^2}{8 \pi}\right]=0   ,
\end{equation}
%%%%%%%%%%%%%%%%%%%%%%%%%%%%%%%%%%%%%%%%%
%%%%%%%%%%%%%%%%%%%%%%%%%%%%%%%%%%%%%%%%%
\begin{equation}\label{eq:basic:momxy}
    \frac{1}{c}\frac{\partial}{\partial t}\left[(\epsilon+p) \gamma^2 \boldsymbol{\beta}_{\boldsymbol{x y}}+\frac{1}{4 \pi}(\boldsymbol{E} \times \boldsymbol{B})_{x y}\right]+\frac{\partial}{\partial z}\left[(\epsilon+p) \gamma^2 \boldsymbol{\beta}_{\boldsymbol{x} \boldsymbol{y}} \beta_z-\frac{1}{4 \pi}\left(\boldsymbol{E}_{\boldsymbol{x y}} E_z+\boldsymbol{B}_{\boldsymbol{x y}} B_z\right)\right]=0,
\end{equation}
%%%%%%%%%%%%%%%%%%%%%%%%%%%%%%%%%%%%%%%%%
%%%%%%%%%%%%%%%%%%%%%%%%%%%%%%%%%%%%%%%%%
\begin{equation}\label{eq:basic:ind_eq}
    \frac{1}{c}\frac{\partial \boldsymbol{B}_{\boldsymbol{x} y}}{\partial t}=[\nabla \times(\boldsymbol{\beta} \times \boldsymbol{B})]_{x y},
\end{equation}
%%%%%%%%%%%%%%%%%%%%%%%%%%%%%%%%%%%%%%%%%
where $\epsilon$ is the energy density (including the rest mass energy density) of matter in the fluid rest frame, $p$ is the pressure, $\beta$ is the 3-velocity normalized by the speed of light $c$, $\gamma=(1-\beta^2)^{-1/2}$ is the Lorentz factor, $\vect{E}$ and $\vect {B}$ are the electric and magnetic fields in the reference frame, respectively.
The subscript $z$ means the component in the $z$ direction and $xy$ means the projection onto the $xy$ plane.
The electric field satisfies the ideal MHD condition\footnote{
\wiadd{
The use of ideal MHD conditions implies the neglect of not only the resistive term but also the Hall term and the electron pressure gradient term in the generalized Ohm's law.
However, this approximation is appropriate for the systems considered in this paper for the following two reasons:
Firstly, the electron pressure gradient and Hall terms are of the order $\mathcal{O}(\omega/\omega_c)$ compared to the Lorentz force,
when the characteristic velocity of the fluid is about Alfv\'en velocity ($\beta\sim\beta_{\rm A}$) and the internal energy of the electron is not too large compared to the kinetic energy associated with the wave motion (e.g., \cite{1992wapl.book.....S,2005ppa..book.....K}).
Therefore, under the current conditions where we are considering sufficiently low-frequency waves, these terms are not significant.
Secondly, in plasmas primarily composed of electron-positron pairs, the positive and negative charges are symmetric, meaning that there is no difference in particle mass or less difference in heating and cooling processes.
Consequently, the Hall term and the electron pressure gradient term are not expected to be major contributors.
}
}:
%%%%%%%%%%%%%%%%%%%%%%%%%%%%%%%%%%%%%%%%%
\begin{equation}\label{eq:basic:idealMHD}
\vect{E}=-\vect{\beta}\times\vect{B}.
\end{equation}
%%%%%%%%%%%%%%%%%%%%%%%%%%%%%%%%%%%%%%%%%
Note that the magnetic field in the $z$ direction should be constant independent of $t$ and $z$, as we can see from the $z$ component of the induction equation and the solenoidal condition $\nabla\cdot\vect{B}=0$.
Moreover, note that a stationary and uniform plasma ($\vect{B}=B_0\vect{e}_z$, $\epsilon={\rm const}$, $p={\rm const}$, and $\vect{\beta}=0$) is a solution to equations (\ref{eq:basic:energy})--(\ref{eq:basic:idealMHD}).
We also assume that the fluid is adiabatic.
As the thermodynamic equation relates between the energy density $\epsilon$ and the pressure $p$, we adopt the following:
%%%%%%%%%%%%%%%%%%%%%%%%%%%%%%%%%%%%%%%%%
\begin{equation}
    \beta_s^2\equiv\left(\pardif{p}{\epsilon}\right)_{\rm ad},
\end{equation}
%%%%%%%%%%%%%%%%%%%%%%%%%%%%%%%%%%%%%%%%%
where $\beta_s$ is the sound speed normalised by $c$ and the subscript ${\rm ad}$ means the adiabatic change.

\subsection{Circularly polarized finite amplitude Alfv\'en wave}
There is an eigenmode in equations (\ref{eq:basic:energy})--(\ref{eq:basic:ind_eq}), which is a circularly polarized Alfv\'en wave.
A circularly polarised Alfv\'en wave is a solution to the MHD equation not only at small amplitudes but also at finite amplitudes.
The circularly polarized Alfv\'en waves propagating in the direction of the background magnetic field $B_0$ are expressed as
%%%%%%%%%%%%%%%%%%%%%%%%%%%%%%%%%%%%%%%%%
\begin{equation}\label{eq:0th:B}
    \boldsymbol{B}=\boldsymbol{B}_0+\delta \boldsymbol{B},
\end{equation}
%%%%%%%%%%%%%%%%%%%%%%%%%%%%%%%%%%%%%%%%%
%%%%%%%%%%%%%%%%%%%%%%%%%%%%%%%%%%%%%%%%%
\begin{equation}\label{eq:0th:Bcomp}
\left\{\begin{aligned}
\delta B_x&=\delta B \cos (k z-\omega t)\\
\delta B_y&=\delta B \sin (k z-\omega t) \\
\delta B_z&=0
\end{aligned}\right.,
\end{equation}
%%%%%%%%%%%%%%%%%%%%%%%%%%%%%%%%%%%%%%%%%
%%%%%%%%%%%%%%%%%%%%%%%%%%%%%%%%%%%%%%%%%
\begin{equation}\label{eq:0th:beta}
\boldsymbol{\beta}=\delta \boldsymbol{\beta}=-\chi_0\beta_{A,\eta} \frac{\delta \boldsymbol{B}}{B_0}, 
\end{equation}
%%%%%%%%%%%%%%%%%%%%%%%%%%%%%%%%%%%%%%%%%
%%%%%%%%%%%%%%%%%%%%%%%%%%%%%%%%%%%%%%%%%
\begin{equation}\label{eq:0th:gas}
\epsilon=\epsilon_0,\quad p=p_0,
\end{equation}
%%%%%%%%%%%%%%%%%%%%%%%%%%%%%%%%%%%%%%%%%
where $\beta_{A,\eta}$ is the phase velocity of the finite amplitude Alfv\'en wave, $\delta \vect{B}$ and $\delta \vect{\beta}$ are the magnetic field and velocity field fluctuations associated with the Alfv\'en wave respectively, and $\chi_0$ is the direction of the wave propagation:
%%%%%%%%%%%%%%%%%%%%%%%%%%%%%%%%%%%%%%%%%
\begin{equation}\label{eq:0th:chi0}
    \chi_0={\rm sgn}(k).
\end{equation}
%%%%%%%%%%%%%%%%%%%%%%%%%%%%%%%%%%%%%%%%%
Furthermore, a dimensionless number representing the amplitude of the wave is introduced as follows:
%%%%%%%%%%%%%%%%%%%%%%%%%%%%%%%%%%%%%%%%%
\begin{equation}
\delta B=\eta B_0.
\end{equation}
%%%%%%%%%%%%%%%%%%%%%%%%%%%%%%%%%%%%%%%%%
Substituting these ansatz into equations (\ref{eq:basic:energy})--(\ref{eq:basic:ind_eq}), we obtain the following from the $x$ component of the induction equation (\ref{eq:basic:ind_eq}):
%%%%%%%%%%%%%%%%%%%%%%%%%%%%%%%%%%%%%%%%%
\begin{equation}\label{eq:0th:dispersionrelation}
\omega=\chi_0\beta_{A,\eta} ck.
\end{equation}
%%%%%%%%%%%%%%%%%%%%%%%%%%%%%%%%%%%%%%%%%
From the $x$ component of the equation of momentum conservation (\ref{eq:basic:momxy}), we obtain
%%%%%%%%%%%%%%%%%%%%%%%%%%%%%%%%%%%%%%%%%
\begin{equation}
\left[\left(\epsilon_0+p_0\right) \delta \gamma^2+\frac{B_0^2}{4 \pi}\right] \chi_0\beta_{A,\eta} \omega-\frac{B_0^2}{4 \pi} c k=0,
\end{equation}
%%%%%%%%%%%%%%%%%%%%%%%%%%%%%%%%%%%%%%%%%
where $\delta\gamma=(1-\delta \beta^2)^{-1/2}=(1-\eta^2 \beta_{A,\eta}^2)^{-1/2}$.
Note that the same equations can be obtained from the $y$ component of the equations (\ref{eq:basic:momxy}) and (\ref{eq:basic:ind_eq}).
By eliminating $\omega$ and $k$ from equations (11) and (12), we obtain the phase velocity $\beta_{A,\eta}$ as follows:
%%%%%%%%%%%%%%%%%%%%%%%%%%%%%%%%%%%%%%%%%
\begin{equation}\label{eq:0th:phasevel}
\beta_{\mathrm{A},\eta}^2=\beta_{\mathrm{A}}^2\left[\frac{1+\beta_{\mathrm{A}}^2 \eta^2}{2}\left(1+\sqrt{1-\left(\frac{2 \beta_{\mathrm{A}}^2 \eta}{1+\beta_{\mathrm{A}}^2 \eta^2}\right)^2}\right)\right]^{-1},
\end{equation}
%%%%%%%%%%%%%%%%%%%%%%%%%%%%%%%%%%%%%%%%%
where $\beta_A$ is a phase velocity of the linear (i.e. infinitesimal amplitude) Alfv\'en wave normalized by the light speed, which can be written as 
%%%%%%%%%%%%%%%%%%%%%%%%%%%%%%%%%%%%%%%%%
\begin{equation}\label{eq:0th:phasevel0}
\beta_{\mathrm{A}}^2 = \frac{B_0^2}{4\pi\left(\epsilon_0+p_0\right)+B_0^2}.
\end{equation}
%%%%%%%%%%%%%%%%%%%%%%%%%%%%%%%%%%%%%%%%%
From equation (\ref{eq:0th:dispersionrelation}), it can be seen that a circularly polarised Alfv\'en wave with finite amplitude expressed in the form of equations (\ref{eq:0th:B})--(\ref{eq:0th:gas}) is a wave propagating with a phase velocity expressed in equation (\ref{eq:0th:phasevel}).
The phase speed of the wave decreases as $\eta$ increases.
For $\eta\ll1$, the phase velocity can be written as:
%%%%%%%%%%%%%%%%%%%%%%%%%%%%%%%%%%%%%%%%%
\begin{equation}\label{eq:fin:phvel}
\beta_{\mathrm{A},\eta} = \beta_{\mathrm{A}}\left[1-\frac{1}{2} \beta_{\mathrm{A}}^2\left(1-\beta_{\mathrm{A}}^2\right) \eta^2+\mathcal{O}\left(\eta^4\right)\right].
\end{equation}
%%%%%%%%%%%%%%%%%%%%%%%%%%%%%%%%%%%%%%%%%
As can be seen from this, the phase speed reduction appears on the order of $\mathcal{O}(\eta^2)$.

\subsection{linearization for daughter waves}\label{sec:geneta_lineps}

%%%%%%%%%%%%%%%%%%%%%%%%%%%%%%%%%%%%%%%%%
\begin{figure}
\includegraphics[width=\linewidth]{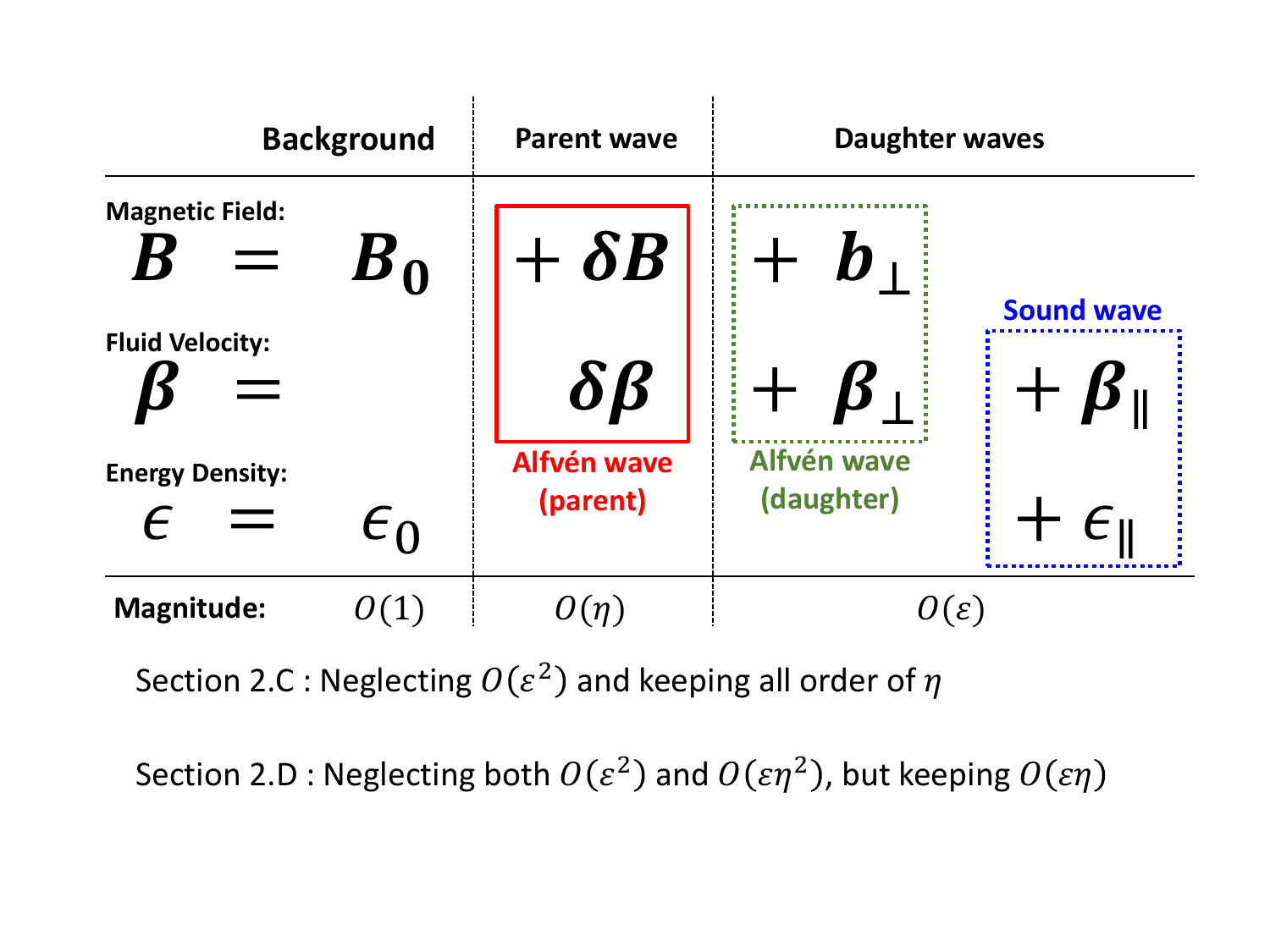}% Here is how to import EPS art
\caption{\label{fig:perturbation}
Summary of setting perturbations in the governing equations.
The terms enclosed by the red rectangle represent the parent Alfv\'en wave.
Note that when combined with background $\mathcal{O}(1)$ quantities (i.e., background), these $\mathcal{O}(\eta)$ \wiadd{($\eta\equiv\delta B/B_0$)} parent waves yield exact solutions to the MHD equations as represented by equations (\ref{eq:0th:B})--(\ref{eq:0th:gas}).
The terms enclosed by the dotted rectangles represent the daughter waves, with the green dotted one representing the daughter Alfv\'en wave and the blue dotted one representing the daughter acoustic waves.
In this paper, we investigate whether there are modes that grow unstable when perturbations represented by dotted lines are added to finite-amplitude circularly polarized Alfv\'en wave.
In Section \ref{sec:geneta_lineps}, $\mathcal{O}(\varepsilon)$ \wiadd{(e.g., $b_\perp/B_0=\mathcal{\varepsilon}$)} is assumed to be small, and equations are derived within its linear range, with no specific assumption on the magnitude of $\eta$.
In Section \ref{sec:lineta_lineps} and later, for simplicity, $\eta$ is considered mildly small, and higher-order terms of $\eta$ are ignored in the perturbation equations in order $\mathcal{O}(\varepsilon)$.
}
\end{figure}
%%%%%%%%%%%%%%%%%%%%%%%%%%%%%%%%%%%%%%%%%

As seen in the previous section, a circularly polarised Alfv\'en wave of finite amplitude is an exact solution of the MHD equation.
In this section, we investigate the stability of Alfv\'en waves by adding perturbations $\vect{b}_\perp$, $\vect{\beta}_\perp$, $\epsilon_{\|}$ and $\beta_{\|}$ of magnitude $\mathcal{O}(\varepsilon)$, expressed in Fig. \ref{fig:perturbation} or in equation (\ref{eq:perturbation}) below, to the solution of the MHD equations (\ref{eq:0th:B})--(\ref{eq:0th:gas}):
%%%%%%%%%%%%%%%%%%%%%%%%%%%%%%%%%%%%%%%%%
\begin{equation}\label{eq:perturbation}
\left\{\begin{array}{rlrrrr}
\boldsymbol{B} & =\boldsymbol{B}_0 & +\delta \boldsymbol{B} & +\boldsymbol{b}_{\perp} & \\
\boldsymbol{\beta} & = & \delta \boldsymbol{\beta} & +\boldsymbol{\beta}_{\perp} & +\boldsymbol{\beta}_{\|} \\
\epsilon & =\epsilon_0 & & & +\epsilon_{\|}
\end{array}\right.,
\end{equation}
%%%%%%%%%%%%%%%%%%%%%%%%%%%%%%%%%%%%%%%%%
where $\vect{b}_\perp$ and $\vect{\beta}_\perp$ are the magnetic field and velocity field fluctuations of the component perpendicular to the $z$ direction (i.e., $\vect{b}_\perp\perp \vect{B}_0$ and $\vect{\beta}_\perp\perp \vect{B}_0$), respectively, meaning Alfv\'en wave-like perturbations.
On the other hand, $\epsilon_{\|}$ and $\beta_{\|}$ are fluctuations of the internal energy density and velocity parallel to the $z$ direction (i.e., $\vect{\beta}_\| \| \vect{B}_0$), respectively, meaning acoustic wave-like perturbations.

Substitute equations (\ref{eq:perturbation}) into equations (\ref{eq:basic:energy})--(\ref{eq:basic:ind_eq}) and equate the coefficients of each order of $\varepsilon$ with $0$.
At the $0$-th order of $\varepsilon$, the non-trivial relations is obtained from equations (\ref{eq:basic:momxy}) and (\ref{eq:basic:ind_eq}) as follows:
%%%%%%%%%%%%%%%%%%%%%%%%%%%%%%%%%%%%%%%%%%%%%%%
\begin{equation}\label{eq:parad:0th:momxy}
	\left[\delta\gamma^2\left(\epsilon_0+p_0\right)+\frac{B_0^2}{4\pi}\right]
	\frac{1}{c}\frac{\partial}{\partial t}\left( \delta\vect{\beta}\right)
	-\frac{B_0}{4 \pi}\pardel{}{z}\left(\delta\vect{B}\right)
	=0,
\end{equation}
%%%%%%%%%%%%%%%%%%%%%%%%%%%%%%%%%%%%%%%%%%%%%%%
%%%%%%%%%%%%%%%%%%%%%%%%%%%%%%%%%%%%%%%%%%%%%%%
\begin{equation}%\label{key}
	\frac{1}{c}\pardel{}{t}\left(\delta\vect{B}\right)-\pardel{}{z}\left(B_0\delta\vect{\beta}\right)=0.
\end{equation}
%%%%%%%%%%%%%%%%%%%%%%%%%%%%%%%%%%%%%%%%%%%%%%%
These equations constitute the wave equation of the parent wave of magnitude $\mathcal{O}(\eta)$.
Note that equations (\ref{eq:basic:energy}) and (\ref{eq:basic:momz}) are trivial at the $0$-th order.
The $1$-st order equations for $\varepsilon$ are as follows (see Appendix \ref{app:derivation} for derivation):
%%%%%%%%%%%%%%%%%%%%%%%%%%%%%%%%%%%%%%%%%
\begin{multline}\label{eq:full:1}
    \frac{1+\beta_s^2 \delta \beta^2}{1-\delta \beta^2} \frac{1}{c} \frac{\partial \epsilon_{\|}}{\partial t}+\left(\delta \gamma^2\left(\epsilon_0+p_0\right)\right) \frac{\partial \beta_{\|}}{\partial z}=\\
    -\frac{1}{c} \frac{\partial}{\partial t}\left[\left(2 \delta \gamma^4\left(\epsilon_0+p_0\right)+\frac{B_0^2}{4 \pi}\right)\left(\delta \boldsymbol{\beta} \cdot \boldsymbol{\beta}_{\perp}\right)+\chi_0 \frac{\delta \beta \delta B B_0 \beta_{\|}}{4 \pi}\right]
    +\frac{B_0}{4 \pi}\left[\boldsymbol{\beta}_{\perp} \cdot \frac{\partial}{\partial z}(\delta \boldsymbol{B})+\delta \boldsymbol{\beta} \cdot \frac{\partial \boldsymbol{b}_{\perp}}{\partial z}\right],
\end{multline}
%%%%%%%%%%%%%%%%%%%%%%%%%%%%%%%%%%%%%%%%%
%%%%%%%%%%%%%%%%%%%%%%%%%%%%%%%%%%%%%%%%%
% \begin{multline}\label{eq:full:2}
\begin{equation}\label{eq:full:2}
\left[\delta \gamma^2\left(\epsilon_0+p_0\right)+\frac{\delta B^2}{4 \pi}\right] \frac{1}{c} \frac{\partial \beta_{\|}}{\partial t}+\beta_s^2 \frac{\partial \epsilon_{\|}}{\partial z}=-\frac{\partial}{\partial z}\left(\frac{\delta \boldsymbol{B} \cdot \boldsymbol{b}_{\perp}}{4 \pi}\right)+\frac{B_0}{4 \pi}\left(\delta \boldsymbol{B} \cdot \frac{1}{c} \frac{\partial \boldsymbol{\beta} \perp}{\partial t}+\boldsymbol{b}_{\perp} \cdot \frac{1}{c} \frac{\partial}{\partial t}(\delta \boldsymbol{\beta})\right),
\end{equation}
% \end{multline}
%%%%%%%%%%%%%%%%%%%%%%%%%%%%%%%%%%%%%%%%%
%%%%%%%%%%%%%%%%%%%%%%%%%%%%%%%%%%%%%%%%%
\begin{multline}\label{eq:full:3}
\left[\delta \gamma^2\left(\epsilon_0+p_0\right)+\frac{B_0^2+\delta B^2}{4 \pi}\right] \frac{1}{c} \frac{\partial \boldsymbol{\beta}_{\perp}}{\partial t}-\frac{B_0}{4 \pi} \frac{\partial \boldsymbol{b}_{\perp}}{\partial z}=-\delta \gamma^2\left(\epsilon_0+p_0\right) \beta_{\|} \frac{\partial}{\partial z}(\delta \boldsymbol{\beta}) \\
-\delta \gamma^2\left(1+\beta_s^2\right) \epsilon_{\|} \frac{1}{c} \frac{\partial}{\partial t}(\delta \boldsymbol{\beta})-\delta \boldsymbol{\beta} \beta_s^2 \frac{1}{c} \frac{\partial \epsilon_{\|}}{\partial t}+\frac{B_0}{4 \pi} \frac{1}{c} \frac{\partial}{\partial t}\left(\beta_{\|} \delta \boldsymbol{B}\right) \\
-2 \delta \gamma^4\left(\epsilon_0+p_0\right)\left(\delta \boldsymbol{\beta} \cdot \boldsymbol{\beta}_{\perp}\right) \frac{1}{c} \frac{\partial}{\partial t}(\delta \boldsymbol{\beta})+\frac{1}{4 \pi} \frac{\partial}{\partial z}\left[\hat{z} \cdot\left\{\left(\delta \boldsymbol{\beta} \times \boldsymbol{b}_{\perp}\right)+\left(\boldsymbol{\beta}_{\perp} \times \delta \boldsymbol{B}\right)\right\}\right]\left(\delta \boldsymbol{\beta} \times \boldsymbol{B}_0\right) \\
\quad-\frac{B_0}{4 \pi} \delta \boldsymbol{\beta}\left[\boldsymbol{\beta}_{\perp} \cdot \frac{\partial}{\partial z}(\delta \boldsymbol{B})+\delta \boldsymbol{\beta} \cdot \frac{\partial \boldsymbol{b}_{\perp}}{\partial z}-B_0 \frac{1}{c} \frac{\partial}{\partial t}\left(\delta \boldsymbol{\beta} \cdot \boldsymbol{\beta}_{\perp}\right)-\chi_0 \delta \beta \delta B \frac{1}{c} \frac{\partial \beta_{\|}}{\partial t}\right] \\
-\frac{\delta \boldsymbol{B} \cdot \boldsymbol{b}_{\perp}}{4 \pi} \frac{1}{c} \frac{\partial}{\partial t}(\delta \boldsymbol{\beta})-\left(\frac{\delta \boldsymbol{B}}{4 \pi} \cdot \frac{1}{c} \frac{\partial \boldsymbol{b}_{\perp}}{\partial t}\right) \delta \boldsymbol{\beta}+\chi_0\frac{\delta B \delta \beta}{4 \pi} \frac{1}{c} \frac{\partial \boldsymbol{b}_{\perp}}{\partial t}+\frac{\delta \boldsymbol{B}}{4 \pi}\left(\frac{1}{c} \frac{\partial  \boldsymbol{\beta}_{\perp}}{\partial t} \cdot \delta \boldsymbol{B}\right)+\frac{\delta \boldsymbol{B} \cdot \boldsymbol{\beta}_{\perp}}{4 \pi} \frac{1}{c} \frac{\partial}{\partial t}(\delta \boldsymbol{B}),
\end{multline}
%%%%%%%%%%%%%%%%%%%%%%%%%%%%%%%%%%%%%%%%%
%%%%%%%%%%%%%%%%%%%%%%%%%%%%%%%%%%%%%%%%%
\begin{equation}\label{eq:full:4}
\frac{1}{c} \frac{\partial \boldsymbol{b}_{\perp}}{\partial t}-B_0 \frac{\partial \boldsymbol{\beta}_{\perp}}{\partial z}=-\frac{\partial}{\partial z}\left(\beta_{\|} \delta \boldsymbol{B}\right).
\end{equation}
%%%%%%%%%%%%%%%%%%%%%%%%%%%%%%%%%%%%%%%%%

\subsection{linearlization for parent wave}\label{sec:lineta_lineps}

Hereafter, for simplicity, we assume a situation where the amplitude of the parent wave is not very large (i.e., $\eta\lesssim 1$), and assume that terms of magnitude $\mathcal{O}(\varepsilon\eta^2)$ are ignored.
In this case, from equation (\ref{eq:fin:phvel}), $\beta_{A,\eta}\sim\beta_A$ for 1-st order equations.
Furthermore, ignoring the term $\mathcal{O}(\varepsilon\eta^2)$ in equations (\ref{eq:full:1})-(\ref{eq:full:4}), we obtain the following equations:
%%%%%%%%%%%%%%%%%%%%%%%%%%%%%%%%%%%%%%%%%
\begin{equation}\label{eq:lin:1}
\frac{1}{c} \frac{\partial \epsilon_{\|}}{\partial t}+\left(\epsilon_0+p_0\right) \frac{\partial \beta_{\|}}{\partial z}=-\frac{1}{c} \frac{\partial}{\partial t}\left[\left(2\left(\epsilon_0+p_0\right)+\frac{B_0^2}{4 \pi}\right)\left(\delta \boldsymbol{\beta} \cdot \boldsymbol{\beta}_{\perp}\right)\right]+\frac{B_0}{4 \pi}\left[\boldsymbol{\beta}_{\perp} \cdot \frac{\partial}{\partial z}(\delta \boldsymbol{B})+\delta \boldsymbol{\beta} \cdot \frac{\partial \boldsymbol{b}_{\perp}}{\partial z}\right],
\end{equation}
%%%%%%%%%%%%%%%%%%%%%%%%%%%%%%%%%%%%%%%%%
%%%%%%%%%%%%%%%%%%%%%%%%%%%%%%%%%%%%%%%%%
\begin{equation}\label{eq:lin:2}
\left(\epsilon_0+p_0\right) \frac{1}{c} \frac{\partial \beta_{\|}}{\partial t}+\beta_s^2 \frac{\partial \epsilon_{\|}}{\partial z}=-\frac{\partial}{\partial z}\left(\frac{\delta \boldsymbol{B} \cdot \boldsymbol{b}_{\perp}}{4 \pi}\right)+\frac{B_0}{4 \pi}\left(\delta \boldsymbol{B} \cdot \frac{1}{c} \frac{\partial \boldsymbol{\beta}_{\perp}}{\partial t}+\boldsymbol{b}_{\perp} \cdot \frac{1}{c} \frac{\partial}{\partial t}(\delta \boldsymbol{\beta})\right),
\end{equation}
%%%%%%%%%%%%%%%%%%%%%%%%%%%%%%%%%%%%%%%%%
%%%%%%%%%%%%%%%%%%%%%%%%%%%%%%%%%%%%%%%%%
\begin{equation}\label{eq:lin:3}
    \mathcal{E} \frac{1}{c} \frac{\partial \boldsymbol{\beta}_{\perp}}{\partial t}-\frac{B_0}{4 \pi} \frac{\partial \boldsymbol{b}_{\perp}}{\partial z}=-\left(\epsilon_0+p_0\right) \beta_{\|} \frac{\partial}{\partial z}(\delta \boldsymbol{\beta})-\left(1+\beta_s^2\right) \frac{B_0 \epsilon_{\|}}{4 \pi \mathcal{E}} \frac{\partial}{\partial z}(\delta \boldsymbol{B})-\delta \boldsymbol{\beta} \beta_s^2 \frac{1}{c} \frac{\partial \epsilon_{\|}}{\partial t}+\frac{B_0}{4 \pi} \frac{1}{c} \frac{\partial}{\partial t}\left(\beta_{\|} \delta \boldsymbol{B}\right),
\end{equation}
%%%%%%%%%%%%%%%%%%%%%%%%%%%%%%%%%%%%%%%%%
%%%%%%%%%%%%%%%%%%%%%%%%%%%%%%%%%%%%%%%%%
\begin{equation}\label{eq:lin:4}
    \frac{1}{c} \frac{\partial \boldsymbol{b}_{\perp}}{\partial t}-B_0 \frac{\partial \boldsymbol{\beta}_{\perp}}{\partial z}=-\frac{\partial}{\partial z}\left(\beta_{\|} \delta \boldsymbol{B}\right),
\end{equation}
%%%%%%%%%%%%%%%%%%%%%%%%%%%%%%%%%%%%%%%%%
where 
%%%%%%%%%%%%%%%%%%%%%%%%%%%%%%%%%%%%%%%%%
\begin{equation}\label{eq:lin:toteneden}
 \mathcal{E} \equiv \epsilon_0+p_0+B_0^2/4 \pi.
\end{equation}
%%%%%%%%%%%%%%%%%%%%%%%%%%%%%%%%%%%%%%%%%
Note that all the terms in the right-hand sides of these equations are $\mathcal{O}(\varepsilon\eta)$, and this implies that $\varepsilon\ll\eta$ is assumed because $\mathcal{O}(\varepsilon^2)$ terms are neglected.
On the right-hand side of these equations, there are some terms that did not appear in the analysis of parametric decay instabilities based on the non-relativistic MHD equations \cite{1969npt..book.....S,1978ApJ...219..700G,1978ApJ...224.1013D}.
% These are terms derived from the velocity squared (i.e., $\mathcal{O}(\beta^2)$), terms in the internal energy density that can be comparable to the rest mass energy density, and terms derived from the displacement current, which were neglected for non-relativistic regime.
These terms stem from (1) the velocity squared (i.e., $\mathcal{O}(\beta^2)$), (2) the internal energy density that can be comparable to the rest mass energy density, and (3) the displacement current, which were neglected for the non-relativistic regime.

\subsection{dimensionless equations}

To clarify the dominated terms in the equation, we normalize the equation and obtain a non-dimensionalized equation.
The quantity representing the amplitude of the parent wave is normalized as follows:
%%%%%%%%%%%%%%%%%%%%%%%%%%%%%%%%%%%%%%%%%
\begin{equation}\label{eq:norm:parent}
\delta \boldsymbol{u} \equiv \frac{\delta \beta}{\beta_A}, \quad \delta \boldsymbol{e}=\frac{\delta \boldsymbol{B}}{B_0}.
\end{equation}
%%%%%%%%%%%%%%%%%%%%%%%%%%%%%%%%%%%%%%%%%
Note that this non-dimensionalisation yields $|\delta \vect{u}|^2=|\delta \vect{e}|^2=\eta^2$.
Also, since $\mathcal{O}(\eta^2)$ is ignored here, the normalization scale of $\delta \vect{\beta}$ is $\beta_{A}$ rather than $\beta_{A,\eta}$.
The amplitude of the daughter Alfv\'en wave of order $\mathcal{O}(\varepsilon)$ is normalized as follows:
%%%%%%%%%%%%%%%%%%%%%%%%%%%%%%%%%%%%%%%%%
\begin{equation}\label{eq:norm:dAlfven}
\boldsymbol{u}_{\perp} \equiv \frac{\beta_{\perp}}{\beta_A}, \quad \boldsymbol{e}_{\perp}=\frac{\boldsymbol{b}_{\perp}}{B_0}.
\end{equation}
%%%%%%%%%%%%%%%%%%%%%%%%%%%%%%%%%%%%%%%%%
For the daughter acoustic wave of order $\mathcal{O}(\varepsilon)$:
%%%%%%%%%%%%%%%%%%%%%%%%%%%%%%%%%%%%%%%%%
\begin{equation}\label{eq:norm:dsound}
u_{\|} \equiv \frac{\beta_{\|}}{\beta_s}, \quad e_{\|} \equiv \frac{\epsilon_{\|}}{\epsilon_0+p_0}.
\end{equation}
%%%%%%%%%%%%%%%%%%%%%%%%%%%%%%%%%%%%%%%%%
% where $w_0\equiv \epsilon_0+p_0$ is the enthalpy density.

Here, we define two dimensionless parameters that characterize this system.
One is the magnetization parameter $\sigma$, defined as follows:
%%%%%%%%%%%%%%%%%%%%%%%%%%%%%%%%%%%%%%%%%
\begin{equation}\label{eq:def:sigma}
\sigma \equiv \frac{B_0^2}{4 \pi\left(\epsilon_0+p_0\right)}.
\end{equation}
%%%%%%%%%%%%%%%%%%%%%%%%%%%%%%%%%%%%%%%%%
By using $\sigma$, the Alfv\'en speed can be written as $\beta_A^2=\sigma/(1+\sigma)$.
The other is the ratio of the sonic velocity to the Alfv\'en velocity:
%%%%%%%%%%%%%%%%%%%%%%%%%%%%%%%%%%%%%%%%%
\begin{equation}\label{eq:def:theta}
\theta \equiv \frac{\beta_s}{\beta_A}.
\end{equation}
%%%%%%%%%%%%%%%%%%%%%%%%%%%%%%%%%%%%%%%%%
Note that this corresponds to approximately the 1/2 power of a quantity called "plasma beta" in the non-relativistic case \footnote{
The plasma beta ($\beta_{\rm pl}$) is typically defined as the ratio of the plasma pressure to the magnetic pressure (i.e., $\beta_{\rm pl}\equiv p_0/(B_0^2/8\pi)$).
The value derived from this definition and the square of the ratio of the sound speed to the Alfv\'en velocity is nearly equal, in fact for a polytropic gas with a specific heat ratio $\Gamma$ (i.e., satisfying $\Gamma$-law), $\theta^2=(2/\Gamma)\beta_{\rm pl}$.
}.

Using equations (\ref{eq:norm:parent})--(\ref{eq:def:theta}), we obtain following normalized equations:
%%%%%%%%%%%%%%%%%%%%%%%%%%%%%%%%%%%%%%%%%
\begin{equation}\label{eq:norm:1st_ene_0}
\frac{1}{c} \frac{\partial e_{\|}}{\partial t}+\beta_s \frac{\partial u_{\|}}{\partial z}=-\beta_A^2 \frac{1}{c} \frac{\partial}{\partial t}\left(\delta \boldsymbol{u} \cdot \boldsymbol{u}_{\perp}\right)-\sigma \delta \boldsymbol{u} \cdot\left(\frac{1}{c} \frac{\partial \boldsymbol{u}_{\perp}}{\partial t}-\beta_A \frac{\partial \boldsymbol{e}_{\perp}}{\partial z}\right),
\end{equation}
%%%%%%%%%%%%%%%%%%%%%%%%%%%%%%%%%%%%%%%%%
%%%%%%%%%%%%%%%%%%%%%%%%%%%%%%%%%%%%%%%%%
\begin{equation}\label{eq:norm:1st_momz_0}
\frac{1}{c} \frac{\partial u_{\|}}{\partial t}+\beta_s \frac{\partial e_{\|}}{\partial z}=-\theta^{-1} \beta_A \frac{\partial}{\partial z}\left(\delta \boldsymbol{e} \cdot \boldsymbol{e}_{\perp}\right)+\sigma \theta^{-1} \delta \boldsymbol{e} \cdot\left(\frac{1}{c} \frac{\partial \boldsymbol{u}_{\perp}}{\partial t}-\beta_A \frac{\partial \boldsymbol{e}_{\perp}}{\partial z}\right),
\end{equation}
%%%%%%%%%%%%%%%%%%%%%%%%%%%%%%%%%%%%%%%%%
%%%%%%%%%%%%%%%%%%%%%%%%%%%%%%%%%%%%%%%%%
\begin{equation}\label{eq:norm:1st_momxy}
\frac{1}{c} \frac{\partial \boldsymbol{u}_{\perp}}{\partial t}-\beta_A \frac{\partial \boldsymbol{e}_{\perp}}{\partial z}=\theta \beta_A^2 \frac{1}{c} \frac{\partial}{\partial t}\left(u_{\|} \delta \boldsymbol{e}\right)-\frac{1}{1+\sigma}\left[\beta_s u_{\|} \frac{\partial}{\partial z}(\delta \boldsymbol{u})+\beta_A e_{\|} \frac{\partial}{\partial z}(\delta \boldsymbol{e})+\beta_s^2 \frac{1}{c} \frac{\partial}{\partial t}\left(e_{\|} \delta \boldsymbol{u}\right)\right],
\end{equation}
%%%%%%%%%%%%%%%%%%%%%%%%%%%%%%%%%%%%%%%%%
%%%%%%%%%%%%%%%%%%%%%%%%%%%%%%%%%%%%%%%%%
\begin{equation}\label{eq:norm:1st_ind}
\frac{1}{c} \frac{\partial \boldsymbol{e}_{\perp}}{\partial t}-\beta_A \frac{\partial \boldsymbol{u}_{\perp}}{\partial z}=-\theta \beta_A \frac{\partial}{\partial z}\left(u_{\|} \delta \boldsymbol{e}\right).
\end{equation}
%%%%%%%%%%%%%%%%%%%%%%%%%%%%%%%%%%%%%%%%%
Here we used the following normalized equation of the order $\mathcal{O}(\varepsilon^0)$:
%%%%%%%%%%%%%%%%%%%%%%%%%%%%%%%%%%%%%%%%%%%%%%%
\begin{equation}\label{eq:norm:0th_eom}
	\frac{1}{c}\frac{\partial}{\partial t}\left(\delta\boldsymbol{u}\right)
	=
	\beta_A\frac{\partial}{\partial z}\left(\delta\boldsymbol{e}\right),
\end{equation}
%%%%%%%%%%%%%%%%%%%%%%%%%%%%%%%%%%%%%%%%%%%%%%%
%%%%%%%%%%%%%%%%%%%%%%%%%%%%%%%%%%%%%%%%%%%%%%%
\begin{equation}\label{eq:norm:0th_ind}
	\frac{1}{c}\frac{\partial}{\partial t}\left(\delta\boldsymbol{e}\right)
	=
	\beta_A\frac{\partial}{\partial z}\left(\delta\boldsymbol{u}\right).
\end{equation}
%%%%%%%%%%%%%%%%%%%%%%%%%%%%%%%%%%%%%%%%%%%%%%%
The second term on the right side of equation (\ref{eq:norm:1st_ene_0}) is found to be a small quantity of $\mathcal{O}(\varepsilon\eta^2)$ using equation (\ref{eq:norm:1st_momxy}).
Therefore, the equation (\ref{eq:norm:1st_ene_0}) becomes
%%%%%%%%%%%%%%%%%%%%%%%%%%%%%%%%%%%%%%%%%
\begin{equation}\label{eq:norm:1st_ene}
\frac{1}{c} \frac{\partial e_{\|}}{\partial t}+\beta_s \frac{\partial u_{\|}}{\partial z}=-\frac{\sigma}{1+\sigma} \frac{1}{c} \frac{\partial}{\partial t}\left(\delta \boldsymbol{u} \cdot \boldsymbol{u}_{\perp}\right).
% -\sigma \delta \boldsymbol{u} \cdot\left(\frac{1}{c} \frac{\partial \boldsymbol{u}_{\perp}}{\partial t}-\beta_A \frac{\partial \boldsymbol{e}_{\perp}}{\partial z}\right)
\end{equation}
%%%%%%%%%%%%%%%%%%%%%%%%%%%%%%%%%%%%%%%%%
Similarly for equation (\ref{eq:norm:1st_momz_0}), we have
%%%%%%%%%%%%%%%%%%%%%%%%%%%%%%%%%%%%%%%%%
\begin{equation}\label{eq:norm:1st_momz}
\frac{1}{c} \frac{\partial u_{\|}}{\partial t}+\beta_s \frac{\partial e_{\|}}{\partial z}=-\theta^{-1} \beta_A \frac{\partial}{\partial z}\left(\delta \boldsymbol{e} \cdot \boldsymbol{e}_{\perp}\right).
% +\sigma \theta^{-1} \delta \boldsymbol{e} \cdot\left(\frac{1}{c} \frac{\partial \boldsymbol{u}_{\perp}}{\partial t}-\beta_A \frac{\partial \boldsymbol{e}_{\perp}}{\partial z}\right)
\end{equation}
%%%%%%%%%%%%%%%%%%%%%%%%%%%%%%%%%%%%%%%%%

\section{Dispersion relation}
As seen in the previous section, the $\mathcal{O}\left(\varepsilon\right)$ perturbation quantities $\vect{u}_{\perp}$, $\vect{e}_{\perp}$, $u_\|$, $e_\|$ follow the linear differential equations (\ref{eq:norm:1st_momxy}), (\ref{eq:norm:1st_ind}), (\ref{eq:norm:1st_ene}), and (\ref{eq:norm:1st_momz}).
In this section, we examine the stability of the wave solutions in these equations by finding the dispersion relation.

\subsection{mode expansion}
In order to find the dispersion relation, each physical quantity of the wave is divided into Fourier components.
Since the parent Alfv\'en wave is a circularly polarized wave with a given wavenumber $k_0$ and frequency $\omega_0$, it can be written as
%%%%%%%%%%%%%%%%%%%%%%%%%%%%%%%%%%%%%%%%%
\begin{equation}\label{eq:mode:parente}
\delta \boldsymbol{e}=\frac{1}{\sqrt{2}}\left(\delta e_0 \exp \left(i \phi_0\right) \vect{e}_{\rm R}+\cc\right),
\end{equation}
%%%%%%%%%%%%%%%%%%%%%%%%%%%%%%%%%%%%%%%%%
%%%%%%%%%%%%%%%%%%%%%%%%%%%%%%%%%%%%%%%%%
\begin{equation}\label{eq:mode:parentu}
\delta \boldsymbol{u}=-\delta \boldsymbol{e},
\end{equation}
%%%%%%%%%%%%%%%%%%%%%%%%%%%%%%%%%%%%%%%%%
where $\phi_0=k_0z-\omega_0t$ is the phase of the parent Alfv\'en wave, $\boldsymbol{e}_{\rm R}=\left(\boldsymbol{e}_x+i \boldsymbol{e}_y\right) / \sqrt{2}$ is the unit polarization vector for right-handed polarization, $\vect{e}_j$ ($j=x,y,z$) is the unit vector, and $\text {c.c.}$ denotes the complex conjugate.
Note that $|\delta e_0|^2=\eta^2$.
Here, although equation (\ref{eq:mode:parente}) seems to represent only right-handed polarization waves, equation (\ref{eq:mode:parente}) simultaneously represents not only right-handed polarized waves with positive $\omega_0$ and positive $k_0$ but also negative $\omega_0$ and negative $k_0$ at the same time \cite{1986JGR....91.4171T}.
Furthermore, in the MHD regime (i.e., where the wave frequency is much smaller than the cyclotron frequency and plasma frequency), right- and left-handed polarizations are symmetric, so that what happens with one polarization also occurs with the other polarization.
Therefore, it is sufficient to discuss wave stability for the parent wave expressed by the equation (\ref{eq:mode:parente}).

For the acoustic wave perturbation, we write the frequency as $\omega$ and the wavenumber as $k$ as follows:
%%%%%%%%%%%%%%%%%%%%%%%%%%%%%%%%%%%%%%%%%
\begin{equation}\label{eq:mode:sounde}
e_{\|}=\frac{1}{2}\left(e_k \exp (i \phi)+\cc\right),
\end{equation}
%%%%%%%%%%%%%%%%%%%%%%%%%%%%%%%%%%%%%%%%%
%%%%%%%%%%%%%%%%%%%%%%%%%%%%%%%%%%%%%%%%%
\begin{equation}\label{eq:mode:soundu}
u_{\|}=\frac{1}{2}\left(u_k \exp (i \phi)+\cc\right),
\end{equation}
%%%%%%%%%%%%%%%%%%%%%%%%%%%%%%%%%%%%%%%%%
where $\phi=kz-\omega t$ is the phase of the acoustic wave.
Since there are two daughter Alfv\'en waves produced by the interaction of the acoustic wave and the parent Alfv\'en wave, corresponding to Stokes and anti-Stokes waves, they are written as follows:
%%%%%%%%%%%%%%%%%%%%%%%%%%%%%%%%%%%%%%%%%
\begin{equation}\label{eq:mode:dalfe}
\boldsymbol{e}_{\perp}=\frac{1}{\sqrt{2}}\left(e_{+} \exp \left(i \phi_{+}\right)\vect{e}_{\rm R}+\cc\right)+\frac{1}{\sqrt{2}}\left(e_{-} \exp \left(i \phi_{-}\right)\vect{e}_{\rm R}+\cc\right),
\end{equation}
%%%%%%%%%%%%%%%%%%%%%%%%%%%%%%%%%%%%%%%%%
%%%%%%%%%%%%%%%%%%%%%%%%%%%%%%%%%%%%%%%%%
\begin{equation}\label{eq:mode:dalfu}
\boldsymbol{u}_{\perp}=\frac{1}{\sqrt{2}}\left(u_{+} \exp \left(i \phi_{+}\right)\vect{e}_{\rm R}+\cc\right)+\frac{1}{\sqrt{2}}\left(u_{-} \exp \left(i \phi_{-}\right)\vect{e}_{\rm R}+\cc\right),
\end{equation}
%%%%%%%%%%%%%%%%%%%%%%%%%%%%%%%%%%%%%%%%%
where $\phi_\pm=\phi_0\pm\phi=k_\pm z-\omega_\pm t$ is the phase of the daughter Alfv\'en waves ($-$ for the Stokes wave, and $+$ for the anti-Stokes wave), and $k_\pm=k_0\pm k$ and $\omega_\pm=\omega_0\pm \omega$ are wavenumbers and frequencies of daughter Alfv\'en waves, respectively.

% (\ref{eq:norm:1st_momxy}), (\ref{eq:norm:1st_ind}), (\ref{eq:norm:1st_ene}), and (\ref{eq:norm:1st_momz}).
Substituting equations (\ref{eq:mode:parente})--(\ref{eq:mode:dalfu}) into equation (\ref{eq:norm:1st_momxy}), the following equation for the Fourier component is obtained from the term proportional to $\exp(i\phi)$:
%%%%%%%%%%%%%%%%%%%%%%%%%%%%%%%%%%%%%%%%%
\begin{equation}\label{eq:parad:hsig_FT:01}
\omega e_k-c\beta_s k u_k=\beta_A^2 \omega\left(u_{+} \delta e_0^*+u_{-}^* \delta e_0\right).
\end{equation}
%%%%%%%%%%%%%%%%%%%%%%%%%%%%%%%%%%%%%%%%%
From equation (\ref{eq:norm:1st_ind}), the following equation is obtained from the $\exp(i\phi)$ component:
%%%%%%%%%%%%%%%%%%%%%%%%%%%%%%%%%%%%%%%%%
\begin{equation}\label{eq:parad:hsig_FT:02}
\omega u_k-c\beta_s k e_k=\theta^{-1} c\beta_A k\left(e_{+} \delta e_0^*+e_{-}^* \delta e_0\right).
\end{equation}
%%%%%%%%%%%%%%%%%%%%%%%%%%%%%%%%%%%%%%%%%
Similarly from the $\exp(i\phi_+)\vect{e}_{\rm R}$ component of equation (\ref{eq:norm:1st_ene}), we obtain the following:
%%%%%%%%%%%%%%%%%%%%%%%%%%%%%%%%%%%%%%%%%
\begin{equation}\label{eq:parad:hsig_FT:03}
\omega_{+} u_{+}+c\beta_A k_{+} e_{+}=\frac{1}{2} \delta e_0\left[\left(\theta \beta_A^2 \omega_{+}-\frac{1}{1+\sigma} \theta c\beta_A k_0\right) u_k
+\frac{1}{1+\sigma}\left(c\beta_A k_0+\theta^2 \beta_A^2 \omega_{+}\right) e_k\right],
\end{equation}
%%%%%%%%%%%%%%%%%%%%%%%%%%%%%%%%%%%%%%%%%
and from the $\exp(-i\phi_-)\vect{e}^*_{\rm R}$ component:
%%%%%%%%%%%%%%%%%%%%%%%%%%%%%%%%%%%%%%%%%
\begin{equation}\label{eq:parad:hsig_FT:04}
\omega_{-} u_{-}+c\beta_A k_{-} e_{-}=\frac{1}{2} \delta e_0^*\left[\left(\theta \beta_A^2 \omega_{-}-\frac{1}{1+\sigma}\theta c\beta_A k_0\right) u_k
+\frac{1}{1+\sigma}\left(c\beta_A k_0+\theta^2 \beta_A^2 \omega_{-}\right) e_k\right].
\end{equation}
%%%%%%%%%%%%%%%%%%%%%%%%%%%%%%%%%%%%%%%%%
Similarly, from the $\exp(i\phi_+)\vect{e}_{\rm R}$ component of equation (\ref{eq:norm:1st_momz}), we obtain
%%%%%%%%%%%%%%%%%%%%%%%%%%%%%%%%%%%%%%%%%
\begin{equation}\label{eq:parad:hsig_FT:05}
\omega_{+} e_{+}+ c\beta_A k_{+} u_{+}=\frac{1}{2} \theta c\beta_A k_{+} u_k \delta e_0,
\end{equation}
%%%%%%%%%%%%%%%%%%%%%%%%%%%%%%%%%%%%%%%%%
and from the $\exp(-i\phi_-)\vect{e}^*_{\rm R}$ component, 
%%%%%%%%%%%%%%%%%%%%%%%%%%%%%%%%%%%%%%%%%
\begin{equation}\label{eq:parad:hsig_FT:06}
\omega_{-} e_{-}^*+ c\beta_A k_{-} u_{-}^*=\frac{1}{2} \theta c\beta_A k_{-} u_k \delta e_0^*.
\end{equation}
%%%%%%%%%%%%%%%%%%%%%%%%%%%%%%%%%%%%%%%%%
Equations (\ref{eq:parad:hsig_FT:01})--(\ref{eq:parad:hsig_FT:06}) are six equations for six Fourier components $e_{k}, u_{k}, e_{+}, u_{+}, e_{-}^{*}, u_{-}^{*}$.

\subsection{dispersion relation}
Hereafter, we adopt the following unit system:
%%%%%%%%%%%%%%%%%%%%%%%%%%%%%%%%%%%%%%%%%
\begin{equation}\label{eq:dr:unit}
\omega_0=1,\quad k_0=1.
\end{equation}
%%%%%%%%%%%%%%%%%%%%%%%%%%%%%%%%%%%%%%%%%
From the conditions for the existence of nontrivial solutions to $e_{k}, u_{k}, e_{+}, u_{+}, e_{-}^{*}, u_{-}^{*}$ in equations (\ref{eq:parad:hsig_FT:01})--(\ref{eq:parad:hsig_FT:06}), the dispersion relation is obtained.
By equating the determinant of the coefficient matrix of equations (\ref{eq:parad:hsig_FT:01})--(\ref{eq:parad:hsig_FT:06}) with $0$, the dispersion relation, which is the equation relating $\omega$ and $k$, is obtained as follows:
% %%%%%%%%%%%%%%%%%%%%%%%%%%%%%%%%%%%%%%%%%
% \begin{eqnarray}
%   (\omega-k)^2 (\omega^2-\theta^2 k^2) \left\{(\omega+k)^2-4\right\}
%   &=&\frac{\eta^2 (\omega-k)}{(1+\sigma)^3}
%   \left(S_0 + S_1 \sigma + S_2 \sigma^2 + S_3 \sigma^3\right)
%   \nonumber\\
%   &+&\frac{\eta^4 k \omega (\omega-k) \sigma}{(1+\sigma)^4}
%   \left(T_0 + T_1 \sigma + T_2 \sigma^2\right)  
% \end{eqnarray}
% %%%%%%%%%%%%%%%%%%%%%%%%%%%%%%%%%%%%%%%%%
%%%%%%%%%%%%%%%%%%%%%%%%%%%%%%%%%%%%%%%%%
\begin{eqnarray}\label{eq:dispersion_relation}
  (\omega-k) (\omega^2-\theta^2 k^2) \left\{(\omega+k)^2-4\right\}
  &=&\frac{\eta^2}{(1+\sigma)^3}
  \left(S_0 + S_1 \sigma + S_2 \sigma^2 + S_3 \sigma^3\right)
  \nonumber\\
  &+&\frac{\eta^4 k \omega\sigma}{(1+\sigma)^4}
  \left(T_0 + T_1 \sigma + T_2 \sigma^2\right),
\end{eqnarray}
%%%%%%%%%%%%%%%%%%%%%%%%%%%%%%%%%%%%%%%%%
where
%%%%%%%%%%%%%%%%%%%%%%%%%%%%%%%%%%%%%%%%%
\begin{eqnarray}
  S_0&=&k^2(\omega^3+k\omega^2-3\omega+k),\\
  S_1&=&2S_0-\omega^3+k\omega^2-\theta^2 k^2
  \left[2k\omega^2+(2k^2-3)\omega-k\right],\\
  S_2&=&S_0-\omega^3+k\omega^2-\theta^2
  \left[\omega^5+2k\omega^4+(k^2-3)\omega^3-3k(k^2+1)\omega^2-k^2(3k^2-7)\omega+k^3\right],\\
  S_3&=&\theta^2 k \omega (\omega-k) \left[(\omega+k)^2-4\right],\\
  T_0&=&k,\\
  T_1&=&-\omega + 2k - \theta^2 (k\omega^2+k^2\omega-k),\\
  T_2&=&-\omega + k - \theta^2 (k\omega-1)(\omega+k).
\end{eqnarray}
%%%%%%%%%%%%%%%%%%%%%%%%%%%%%%%%%%%%%%%%%
Note that since we have only calculated correctly up to the lowest order of $\eta$ in this paper, expressions of $T_j$ ($\mathcal{O}(\eta^2)$ times higher order terms than $S_j~(j=0,1,2,3)$) may change when the calculation correctly incorporates higher order terms of $\eta$.
Note also that although we used the dispersion relation $\omega_0=c\beta_A k_0$ for the parent Alfv\'en wave propagating in the positive $z$ axis in deriving equation (\ref{eq:dispersion_relation}), the above expression is identical for a wave propagating in the opposite direction (i.e. the negative $z$ direction).
Equation (\ref{eq:dispersion_relation}) is an algebraic equation for $\omega$ and $k$, and its solution represents the wave that can propagate through the plasma in the presence of parent Alfv\'en wave.

%%%%%%%%%%%%%%%%%%%%%%%%%%%%%%%%%%%%%%%%%
\begin{figure}[t]
\includegraphics{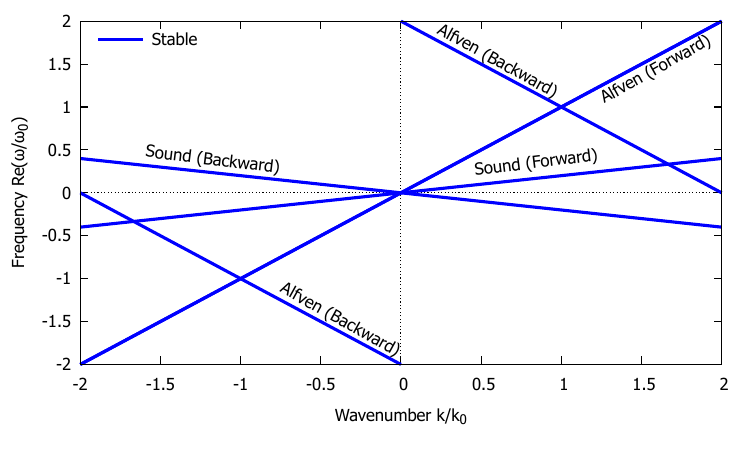}% Here is how to import EPS art
\caption{\label{fig:eta0} 
Dispersion relation (\ref{eq:dispersion_relation}) with $\eta=0$, and $\theta=0.2$.
The horizontal axis is the wavenumber normalized by that of the parent Alfv\'en wave $k_0$ and the vertical axis is the frequency normalized by $\omega_0$.
Each line corresponds to the $\mathcal{O}(\varepsilon)$ wave propagating in a uniform plasma in a uniform magnetic field.
The backward propagating Alfv\'en wave in the first quadrant represents the daughter Alfv\'en wave (Stokes wave) with frequency $\omega_-=\omega_0-\omega$ and wavenumber $k_-=k_0-k$.
The backward Alfv\'en wave in the third quadrant is the daughter Alfv\'en wave (anti-Stokes wave) with frequency $\omega_+=\omega_0+\omega$ and wavenumber $k_+=k_0+k$.
}
\end{figure}
%%%%%%%%%%%%%%%%%%%%%%%%%%%%%%%%%%%%%%%%%

For $\eta=0$, equation (\ref{eq:dispersion_relation}) recovers the dispersion relation for waves propagating parallel to the magnetic field in a uniform plasma.
Figure \ref{fig:eta0} shows the solution of equation (\ref{eq:dispersion_relation}) for $\eta=0$ and $\theta=0.2$.
Since the parent Alfv\'en wave is propagating in the $z$-axis positive direction, $\omega-k=0$ can be written as $\omega_\pm=c\beta_A k_\pm$ (in the physical unit), which represents the daughter Alfv\'en waves (Stokes and anti-Stokes waves) propagating in the same direction as the parent wave.
From $\omega+k-2=0$, this can be written in terms of the frequency and wavenumber of the daughter Alfv\'en wave as $\omega_-=-c\beta_Ak_-$, indicating that it is the daughter Alfv\'en wave (Stokes wave) propagating in the opposite direction of the parent wave.
Similarly, $\omega+k+2=0$ becomes $\omega_+=-c\beta_Ak_+$, which represents the daughter Alfv\'en wave (anti-Stokes wave) propagating in the opposite direction to the parent wave.
Furthermore, $\omega^2-\theta^2k^2=0$ represents acoustic waves propagating backward or forward in the $z$ direction, as shown by the fact that $\omega^2=\beta_s^2k^2$ in the physical unit.

%%%%%%%%%%%%%%%%%%%%%%%%%%%%%%%%%%%%%%%%%
\begin{figure}[b]
\includegraphics{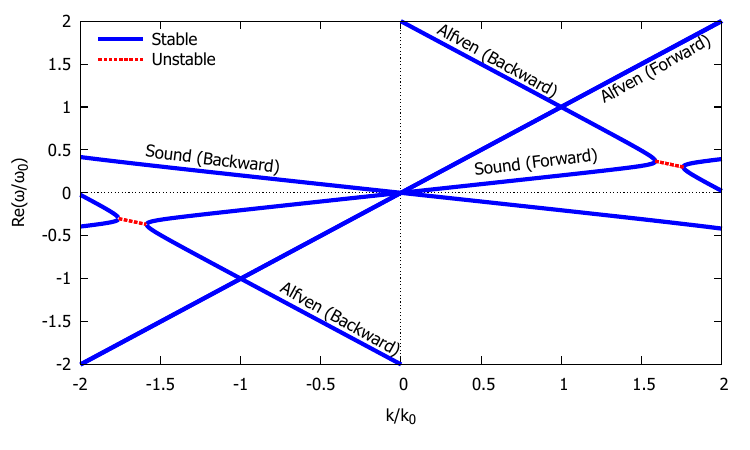}% Here is how to import EPS art
\caption{\label{fig:disp}
Dispersion relation (\ref{eq:dispersion_relation}) for $\eta=0.2$, $\theta=0.2$, and $\sigma=10$.
Same as Figure \ref{fig:eta0}, the horizontal axis is the wavenumber normalized by $k_0$ and the vertical axis is the real part of the frequency normalized by $\omega_0$.
The solid blue curves represent the solutions for stable waves, and the dotted red lines represent the solutions for unstable growing waves.
% It can be seen that waves of frequencies and wavenumbers that satisfy the dispersion relation for both forward acoustic waves and backward Alfv\'en waves are unstable.
It can be seen that the instability occurs at the intersection point ($k\sim \pm1.7k_0$, $\omega\sim\pm0.3\omega_0$) for the forward acoustic wave and the backward daughter Alfv\'en wave (Stokes waves).
}
\end{figure}
%%%%%%%%%%%%%%%%%%%%%%%%%%%%%%%%%%%%%%%%%

%%%%%%%%%%%%%%%%%%%%%%%%%%%%%%%%%%%%%%%%%
\begin{figure}[b]
\includegraphics{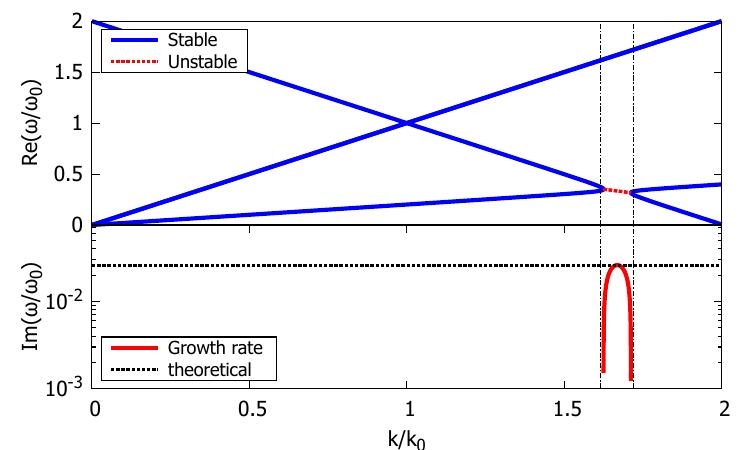}% Here is how to import EPS art
\caption{\label{fig:disp_rate1}
Dispersion relation (\ref{eq:dispersion_relation}) with $\eta=0.2$, $\theta=0.2$, and $\sigma=10$.
The upper panel shows the real part of the frequency $\omega$ (the same figure as the first quadrant in Figure \ref{fig:disp}) and the lower panel shows the imaginary part of the frequency $\omega$.
The horizontal axis is the wavenumber normalized by $k_0$.
The horizontal line in the lower panel, indicated by the black thick dotted line, is the maximum growth rate $\Gamma_{\rm max}$, obtained by the equation (\ref{eq:growthrate_gen}), which is in good agreement with the maximum value of the growth rate obtained by numerical calculation.
The vertical dashed-dotted lines in the upper and lower panels represent the upper and lower limits of the range of wavenumbers $k$ where instability occurs, as expressed in equation (\ref{eq:deltak_gen}).
}
\end{figure}
%%%%%%%%%%%%%%%%%%%%%%%%%%%%%%%%%%%%%%%%%

%%%%%%%%%%%%%%%%%%%%%%%%%%%%%%%%%%%%%%%%%
\begin{figure}[b]
\includegraphics{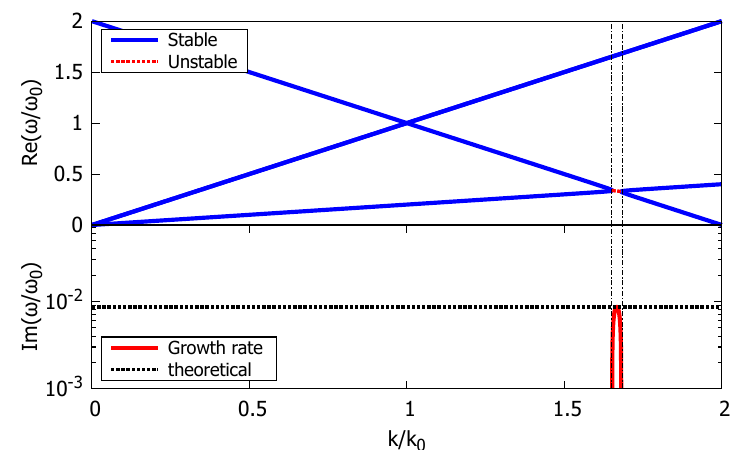}% Here is how to import EPS art
\caption{\label{fig:disp_rate2}
Dispersion relation (\ref{eq:dispersion_relation}) calculated with $\sigma=100$, while $\eta$ and $\theta$ are the same as in Figure \ref{fig:disp_rate1}.
}
\end{figure}
%%%%%%%%%%%%%%%%%%%%%%%%%%%%%%%%%%%%%%%%%

Figure \ref{fig:disp} shows the dispersion relation obtained by solving the equation (\ref{eq:dispersion_relation}) under the parameters $\sigma=10$, $\eta=0.2$, and $\theta=0.2$.
The curves shown in blue in the figure represent solutions with stable waves, while the curves shown in red dotted lines represent solutions that have non-zero imaginary parts (i.e., solutions with unstable growth).
% The instability occurs in waves of frequencies and wavenumbers that satisfy the dispersion relation for both forward acoustic waves and backward daughter Alfv\'en waves (Stokes waves).
In the figure, the instability occurs at the intersection point of the dispersion relations for the forward acoustic wave and the backward daughter Alfv\'en wave (Stokes waves).
Figure \ref{fig:disp_rate1} shows the real (upper panel) and imaginary (lower panel) parts of the frequency when the dispersion relation (\ref{eq:dispersion_relation}) is solved with the same parameters as in Figure \ref{fig:disp}.
As can be seen from these, even for plasma with relativistic magnetization (in this case $\sigma=10$), the three-wave interaction which satisfies the resonance conditions $\omega_0=\omega+\omega_-$ and $k_0=k+k_-$ is unstable.

\subsection{Non-relativistic limit}
In order to check the validity of the obtained dispersion relation, we take the non-relativistic limit and compare it with the known results.
The non-relativistic limit here means the limit where the velocity scale of the system is much slower than the speed of light and the rest mass energy density is much larger than the internal energy density.
This can be achieved by taking the limit of Alfv\'en velocity $\beta^2_A\ll1$, or since $\sigma=\beta_A^2/(1-\beta_A^2)$, the limit of $\sigma\to0$.
By taking the limit in equation (\ref{eq:dispersion_relation}), we obtain
%%%%%%%%%%%%%%%%%%%%%%%%%%%%%%%%%%%%%%%%%
\begin{equation}
% (\omega-k)^2\left(\omega^2-\theta^2 k^2\right)\left\{(\omega+k)^2-4\right\}=\eta^2 k^2(\omega-k)\left(\omega^3+k \omega^2-3 \omega+k\right).
(\omega-k)\left(\omega^2-\theta^2 k^2\right)\left\{(\omega+k)^2-4\right\}=\eta^2 k^2\left(\omega^3+k \omega^2-3 \omega+k\right).
\end{equation}
%%%%%%%%%%%%%%%%%%%%%%%%%%%%%%%%%%%%%%%%%
This is the same dispersion relation as previous results obtained for non-relativistic MHD equation system \cite{1978ApJ...219..700G,1978ApJ...224.1013D}.
Therefore, in this regime ($\sigma\ll1$), based on this dispersion relation, the daughter Alfv\'en and the acoustic waves grow exponentially.

\subsection{$\sigma\to\infty$ limit}
As another extreme case, let us consider the case where $\sigma\to\infty$.
The dispersion relation (\ref{eq:dispersion_relation}) leaving only the highest order of $\sigma$ is as follows:
%%%%%%%%%%%%%%%%%%%%%%%%%%%%%%%%%%%%%%%%%
\begin{equation}
% (\omega-k)^2\left\{(\omega+k)^2-4\right\}\left(\omega^2-\eta^2 \theta^2 k \omega-\theta^2 k^2\right)=0.
(\omega-k)\left\{(\omega+k)^2-4\right\}\left(\omega^2-\eta^2 \theta^2 k \omega-\theta^2 k^2\right)=0.
\end{equation}
%%%%%%%%%%%%%%%%%%%%%%%%%%%%%%%%%%%%%%%%%
Since there are only real solutions of $\omega$ to this equation for any $k$, $\theta$, and $\eta$, it can be seen that instability does not occur in the limit of $\sigma\to\infty$.

\subsection{$1/\sigma$ expansion}\label{sec:disp:exp}
As seen in the previous two sections, in the limit of small $\sigma$ (i.e., non-relativistic limit), the PDI occurs, but in the limit of $\sigma\to\infty$, no instability occurs.
This fact implies that the parametric decay instability (PDI) is suppressed as $\sigma$ increases.
In order to investigate this quantitatively, we expand the dispersion relation with $\sigma^{-1}$ as a small quantity and investigate how the instability is suppressed up to the first order of $\sigma^{-1}$.
Hereafter, the term $\eta^4$ in the dispersion relation is ignored as small, and $\theta\ll1$ is assumed for simplicity.
The dispersion relation (\ref{eq:dispersion_relation}) expanded to the first order of $\sigma^{-1}$ is as follows:
%%%%%%%%%%%%%%%%%%%%%%%%%%%%%%%%%%%%%%%%%
\begin{equation}\label{eq:exp:disp_sig}
% (\omega-k)^2\left(\omega^2-\theta^2 k^2\right)\left[(\omega+k)^2-4\right]
% =
% \eta^2(\omega-k)\left[S_3+\frac{1}{\sigma}\left(S_2-3 S_3\right)\right].
(\omega-k)\left(\omega^2-\theta^2 k^2\right)\left[(\omega+k)^2-4\right]
=
\eta^2\left[S_3+\frac{1}{\sigma}\left(S_2-3 S_3\right)\right].
\end{equation}
%%%%%%%%%%%%%%%%%%%%%%%%%%%%%%%%%%%%%%%%%
Since this equation is a $ 6$-th-order algebraic equation, it cannot be solved analytically in general.
Therefore, in order to derive the conditions for the existence of unstable solutions, we set $\eta$ to be small and expand the dispersion relation around the following resonance frequency $\omega_1$ and wavenumber $k_1$, which satisfies both the dispersion relation $\omega=\theta k$ for forward acoustic waves and $\omega+k-2=0$ for backward Alfv\'en waves (Stokes waves).
The resonance frequency $\omega_1$ and wavenumber $k_1$ can be written as
%%%%%%%%%%%%%%%%%%%%%%%%%%%%%%%%%%%%%%%%%
\begin{equation}\label{eq:disp:exp:resonance}
    \omega_1=\frac{2\theta}{1+\theta},~~ k_1=\frac{2}{1+\theta}.
\end{equation}
%%%%%%%%%%%%%%%%%%%%%%%%%%%%%%%%%%%%%%%%%
We take the frequency $\omega$ and the wavenumber $k$ as follows:
%%%%%%%%%%%%%%%%%%%%%%%%%%%%%%%%%%%%%%%%%
\begin{equation}\label{eq:dispersion:exp:vars}
    \omega=\omega_1+\delta\omega,~~k=k_1+\delta k.
\end{equation}
%%%%%%%%%%%%%%%%%%%%%%%%%%%%%%%%%%%%%%%%%
We assume that $\delta\omega$ and $\delta k$ are small quantities of magnitude about $\mathcal{O}(\eta)$.
This will be confirmed later to be correct in the case where the resonance condition is satisfied.
Expanding the left-hand side of equation (\ref{eq:exp:disp_sig}) to the lowest order of $\eta$, we obtain
%%%%%%%%%%%%%%%%%%%%%%%%%%%%%%%%%%%%%%%%%
\begin{equation}
(\omega-k)\left(\omega^2-\theta^2 k^2\right)\left\{(\omega+k)^2-4\right\}
\sim -32 \theta\delta\omega\left(\delta \omega+\delta k\right).
\end{equation}
%%%%%%%%%%%%%%%%%%%%%%%%%%%%%%%%%%%%%%%%%
Note that here we used $\theta\ll1$ and keep only the lowest order of $\theta$, but assuming that $\theta$ is greater than $\mathcal{O}(\eta)$.
Similarly, the following equation can be obtained for the right-hand side by expanding to the lowest order of $\eta$:
%%%%%%%%%%%%%%%%%%%%%%%%%%%%%%%%%%%%%%%%%
\begin{equation}
\eta^2\left[S_3+\frac{1}{\sigma}\left(S_2-3 S_3\right)\right]
\sim
\frac{8\eta^2}{\sigma}.
\end{equation}
%%%%%%%%%%%%%%%%%%%%%%%%%%%%%%%%%%%%%%%%%
Equating the above results, we thus obtain:
%%%%%%%%%%%%%%%%%%%%%%%%%%%%%%%%%%%%%%%%%
\begin{equation}\label{eq:exp:perturbedeq}
\delta\omega^2+\delta k\delta\omega + \frac{\eta^2}{4\sigma\theta}=0.
\end{equation}
%%%%%%%%%%%%%%%%%%%%%%%%%%%%%%%%%%%%%%%%%
If equation (\ref{eq:exp:perturbedeq}) has a solution with an imaginary part, equation (\ref{eq:exp:disp_sig}) has an unstable growing solution.
Regarding equation (\ref{eq:exp:perturbedeq}) as a quadratic equation of $\delta \omega$, the condition that its discriminant is negative gives the following condition that $\delta \omega$ has an imaginary part, namely:
%%%%%%%%%%%%%%%%%%%%%%%%%%%%%%%%%%%%%%%%%
\begin{equation}\label{eq:exp:condition:deltak}
-\eta\sigma^{-1/2}\theta^{-1/2}<\delta k<\eta\sigma^{-1/2}\theta^{-1/2}.
\end{equation}
%%%%%%%%%%%%%%%%%%%%%%%%%%%%%%%%%%%%%%%%%
The imaginary part $\Gamma\equiv{\rm Im}(\delta \omega)$ takes its maximum value $\Gamma_{\rm max}$ when the resonance condition is satisfied (i.e. $\delta k=0$) and is as follows:
%%%%%%%%%%%%%%%%%%%%%%%%%%%%%%%%%%%%%%%%%
\begin{equation}\label{eq:growthrate_sig}
\Gamma_{\rm max}=\frac{1}{2}\eta\sigma^{-1/2}\theta^{-1/2}.
\end{equation}
%%%%%%%%%%%%%%%%%%%%%%%%%%%%%%%%%%%%%%%%
Note that the validity of the assumption that $\delta \omega$ and $\delta k$ are $\mathcal{O}(\eta)$ can be confirmed from equations (\ref{eq:exp:condition:deltak}) and (\ref{eq:growthrate_sig}).
As can be seen from equations (\ref{eq:exp:condition:deltak}) and (\ref{eq:growthrate_sig}), the instability ceases in the limit of $\sigma\to\infty$.

\subsection{Decay rate for general $\sigma$}\label{sec:gensig}
By using the same method used in section \ref{sec:disp:exp} to equation (\ref{eq:dispersion_relation}), we can obtain the growth rate and the conditions under which instability occurs for general $\sigma$.
In this section, we continue to use $\mathcal{O}(\eta)$ as a small quantity and ignore the term of $T_j$ on the right-hand side of the equation (\ref{eq:dispersion_relation}).
Substituting equation (\ref{eq:dispersion:exp:vars}) into equation (\ref{eq:dispersion_relation}) and deriving the lowest order equation for $\eta$, we obtain:
%%%%%%%%%%%%%%%%%%%%%%%%%%%%%%%%%%%%%%%%%
\begin{equation}
\delta\omega^2
+\left(1-\theta\right)\delta k\delta \omega
-\theta \delta k^2
+\frac{\eta^2\left(1-\theta\right)}{4\theta\left(1+\theta\right)^2}\frac{\left[1+\sigma\left(1+\theta^2\right)\right]^2}{\left(1+\sigma\right)^3}
=0.
\end{equation}
%%%%%%%%%%%%%%%%%%%%%%%%%%%%%%%%%%%%%%%%%
The condition for $\delta\omega$ to have a solution with an imaginary part is as follows:
%%%%%%%%%%%%%%%%%%%%%%%%%%%%%%%%%%%%%%%%%
\begin{equation}\label{eq:deltak_gen}
-\eta\theta^{-1/2}\frac{\sqrt{1-\theta}}{\left(1+\theta\right)^2}\frac{1+\sigma\left(1+\theta^2\right)}{\left(1+\sigma\right)^{3/2}}
<
\delta k
<
\eta\theta^{-1/2}\frac{\sqrt{1-\theta}}{\left(1+\theta\right)^2}\frac{1+\sigma\left(1+\theta^2\right)}{\left(1+\sigma\right)^{3/2}}.
\end{equation}
%%%%%%%%%%%%%%%%%%%%%%%%%%%%%%%%%%%%%%%%%
The growth rate is at maximum $\Gamma_{\rm max}$ when $\delta k=0$, and it can be written as:
%%%%%%%%%%%%%%%%%%%%%%%%%%%%%%%%%%%%%%%%%
\begin{equation}\label{eq:growthrate_gen}
\Gamma_{\rm max}=\frac{1}{2}\eta\theta^{-1/2}\frac{\sqrt{1-\theta}}{1+\theta}\frac{1+\sigma\left(1+\theta^2\right)}{\left(1+\sigma\right)^{3/2}}.
\end{equation}
%%%%%%%%%%%%%%%%%%%%%%%%%%%%%%%%%%%%%%%%%
In the non-relativistic limit (i.e. $\sigma\ll1$) and the low-beta limit (i.e. $\theta\ll1$), this expression becomes $\Gamma_{\rm max}=\eta \theta^{-1/2}/2$ and reproduces the well-known expression of the growth rate of the PDI for non-relativistic plasma.

This formula (\ref{eq:growthrate_gen}) for the growth rate agrees with Matsukiyo \& Hada (2003) \cite{2003PhRvE..67d6406M} within the range of $\mathcal{O}(\theta)$ \footnote{Although Matsukiyo \& Hada (2003) assumed $\sigma\lesssim1$ when deriving the growth rate, the $\sigma$-dependence of their formula is consistent with our result for general $\sigma$.
They justified neglecting some terms when expanding the dispersion relation around the resonance point by assuming that $\sigma$ is not too large.
However, considering only points well close to the resonance point is sufficient to neglect these terms, making the assumption about $\sigma$ unnecessary.
}.
The discrepancy beyond $\mathcal{O}(\theta)$ arises because their formulation implicitly assumes that the internal energy is significantly smaller than the rest mass energy.
Furthermore, the above result is also approximately derived from the growth rate of the induced Brillouin scattering for unmagnetized pair plasma, equation (A3) in Iwamoto et al. (2023) \cite{2023MNRAS.522.2133I}, by multiplying $(\omega/\omega_{\rm c})^{2}$, where $\omega_{\rm c}$ is the electron cyclotron frequency.
This is because the oscillation amplitude of electrons perpendicular to $\vect{B}_{0}$ is suppressed by $\omega/\omega_{\rm c}$ in the presence of a background magnetic field compared to the scattering of electromagnetic waves without a magnetic field.

In the bottom panel of Figure \ref{fig:disp_rate1}, the maximum growth rate obtained from equation (\ref{eq:growthrate_gen}) is shown by a horizontal thick black dotted line.
The upper and lower limits of the range of wavenumber $k$ in which instability occurs, represented by the equation (\ref{eq:deltak_gen}), are also indicated by the vertical dashed-dotted lines in the upper and lower panels.
As is clear from the figure, the obtained analytical expressions reproduce the numerical results well.
Figure \ref{fig:disp_rate2} shows the dispersion relation when $\sigma=10^2$ and $\eta$ and $\theta$ are the same as in figure \ref{fig:disp_rate1}.
As $\sigma$ increases, not only the maximum growth rate becomes smaller, but also the range of instability becomes narrower.
Takamoto et al. (2014) \cite{2014ApJ...787...84T} performed simulations of Alfv\'en wave propagation based on the relativistic MHD equation and obtained a qualitative trend that Alfv\'en waves tend to be more stable as $\sigma$ increases.
Our result, equation (\ref{eq:growthrate_gen}), is consistent with their results and gives a quantitative interpretation of their nature.

%%%%%%%%%%%%%%%%%%%%%%%%%%%%%%%%%%%%%%%%%
\begin{figure}
\includegraphics[width=\linewidth]{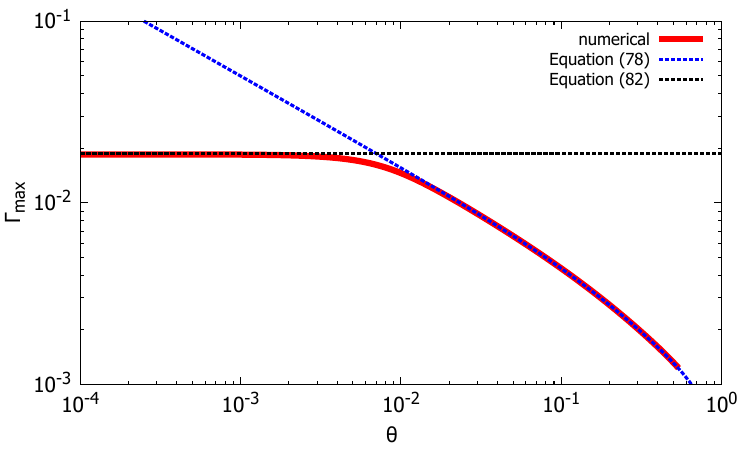}% Here is how to import EPS art
\caption{\label{fig:theta_dep}
$\theta$-dependence of the maximum growth rate $\Gamma_{\rm max}$ for $\sigma=10^3$ and $\eta=0.2$.
Under these parameters, $\theta_c\sim0.01$ (see equation (\ref{eq:small:theta_c})).
The red curve is the numerical solution of the dispersion relation (\ref{eq:dispersion_relation}) and the blue dotted curve is the growth rate obtained from the equation (\ref{eq:growthrate_gen}).
For $\theta\gtrsim\theta_c$, these are in good agreement.
The black dotted line is the growth rate in the limit where $\theta\lesssim\theta_c$, calculated by equation (\ref{eq:growthrate_smalltheta}).
When $\theta$ is small, the growth rate takes a constant value independent of $\theta$.
}
\end{figure}
%%%%%%%%%%%%%%%%%%%%%%%%%%%%%%%%%%%%%%%%%

Figure \ref{fig:theta_dep} shows the dependence of the maximum growth rate $\Gamma_{\rm max}$ on $\theta$ when $\sigma=10^3$ and $\eta=0.2$.
The red curve is the result of numerically and directly solving the dispersion relation (\ref{eq:dispersion_relation}), and the blue dotted curve is the growth rate calculated from equation (\ref{eq:growthrate_gen}).
It can be seen that the analytical formula (\ref{eq:growthrate_gen}) is in good agreement with the numerical solution in $\theta\gtrsim10^{-2}$.
However, for $\theta\lesssim10^{-2}$, equation (\ref{eq:growthrate_gen}) fails to reproduce the numerical result.
This is because the approximation of $|\delta\omega|\ll\omega_1$ is broken.
The correct growth rate in such a small limit of $\theta$ will be obtained in the next section.

\subsection{Strong coupling limit: decay rate for small $\theta$}
As seen in Figure \ref{fig:theta_dep}, when $\theta$ is small, equation (\ref{eq:growthrate_gen}) fails to reproduce the numerical solution.
This is due to the fact that the condition $|\delta\omega|\ll\omega_1$, which was assumed in obtaining equation (\ref{eq:growthrate_gen}), is violated.
In the context of plasma physics, such a situation is called strong coupling limit \cite{1975PhFl...18.1002F,1979PhFl...22.1115C}.
The typical $\theta=\theta_c$ where the approximation begins to break down is estimated from $\omega_1\sim|\delta\omega|$ using equations (\ref{eq:disp:exp:resonance}) and (\ref{eq:growthrate_gen}) as follows:
%%%%%%%%%%%%%%%%%%%%%%%%%%%%%%%%%%%%%%%%%
\begin{equation}\label{eq:small:theta_c}
\theta_c \sim 2^{-4 / 3} \eta^{2 / 3}\left(1+\sigma\right)^{-1 / 3}.
\end{equation}
%%%%%%%%%%%%%%%%%%%%%%%%%%%%%%%%%%%%%%%%%
Since $\Gamma_{\rm max}\sim \Gamma_{\rm max}(\theta=\theta_c)$ at $\theta\ll\theta_c$ (see Figure \ref{fig:theta_dep}), we can presume that the growth rate at $\theta\ll\theta_c$ is proportional to $\eta^{2/3}$ from equation (\ref{eq:growthrate_gen}).
Therefore, we assume that $\delta\omega$ and $\delta k$ are $\mathcal{O}(\eta^{2/3})$ and expand the dispersion relation again.
Note that $\theta$ in this regime is smaller than $\mathcal{O}(\eta^{2/3})$ because $\theta_c=\mathcal{O}(\eta^{2/3})$.
Expanding the equation (\ref{eq:dispersion_relation}), we obtain the following equation from the lowest order $\mathcal{O}(\eta^2)$ terms:
%%%%%%%%%%%%%%%%%%%%%%%%%%%%%%%%%%%%%%%%%
\begin{equation}\label{eq:exp:lowtheta:eqs_pert}
\delta\omega^2\left(\delta\omega+\delta k\right)+\frac{\eta^2}{1+\sigma}=0.
\end{equation}
%%%%%%%%%%%%%%%%%%%%%%%%%%%%%%%%%%%%%%%%%
The condition that the solution $\delta\omega$ of equation (\ref{eq:exp:lowtheta:eqs_pert}) has an imaginary part is that the wavenumber $k$ is greater than the following value $k_{\rm min}$:
%%%%%%%%%%%%%%%%%%%%%%%%%%%%%%%%%%%%%%%%%
\begin{equation}
k>k_{\rm min}=2-\sqrt[3]{\frac{27}{4}}\eta^{2/3}\left(1+\sigma\right)^{-1/3}.
\end{equation}
%%%%%%%%%%%%%%%%%%%%%%%%%%%%%%%%%%%%%%%%%
When $\delta k=0$, the growth rate takes the maximum value $\Gamma_{\rm max}$ as follows:
%%%%%%%%%%%%%%%%%%%%%%%%%%%%%%%%%%%%%%%%%
\begin{equation}\label{eq:growthrate_smalltheta}
\Gamma_{\rm max}=\frac{\sqrt{3}}{2} \eta^{2 / 3} \left(1+\sigma\right)^{-1 / 3}.
\end{equation}
%%%%%%%%%%%%%%%%%%%%%%%%%%%%%%%%%%%%%%%%%

The black dotted line in Figure \ref{fig:theta_dep} represents the growth rate calculated by equation (\ref{eq:growthrate_smalltheta}).
In the parameters of Figure \ref{fig:theta_dep}, $\theta_c\sim0.01$, it can be seen that equation (\ref{eq:growthrate_smalltheta}) reproduces well the growth rate obtained by the numerical calculation for $\theta\ll\theta_c$.
Note that, in the non-relativistic limit (i.e. $\sigma\to0$), it reproduces the growth rate of PDI in non-relativistic plasmas at $\theta\sim0$ \cite{1993JGR....9819049J}.

\section{Discussion}

The obtained expression for the growth rate of the instability reveals that the maximum growth rate $\Gamma_{\rm max}$ is a decreasing function of $\sigma$, and in the limit of $\sigma\to\infty$, the instability ceases.
Since $\sigma$ is the ratio of the energy densities of the matter and electromagnetic fields, the limit of large $\sigma$ corresponds to a situation close to the force-free limit.
Our findings are consistent with the fact that an Alfv\'en wave is stable to resonant three-wave interactions in the force-free limit.
While it is not evident whether the methods using force-free equations and the MHD equations employed in our study are mathematically equivalent, we can demonstrate that they are equivalent within the scope of this study.
By eliminating $\vect{u}_\perp$ from equations (\ref{eq:norm:1st_momxy}) and (\ref{eq:norm:1st_ind}), and taking the limit of $\sigma\to \infty$, we obtain
%%%%%%%%%%%%%%%%%%%%%%%%%%%%%%%%%%%%%%%%%
\begin{equation}\label{eq:discussion:ff1}
\frac{1}{c^2}\frac{\partial^2 \boldsymbol{b}_{\perp}}{\partial t^2}-\frac{\partial^2 \boldsymbol{b}_{\perp}}{\partial z^2}=0.
\end{equation}
%%%%%%%%%%%%%%%%%%%%%%%%%%%%%%%%%%%%%%%%%
This is the equation satisfied by the magnetic field of the free electromagnetic wave.
Similarly, by eliminating $\vect{e}_\perp$ from equations (\ref{eq:norm:1st_momxy}) and (\ref{eq:norm:1st_ind}), and taking the limit of $\sigma\to \infty$, we obtain
%%%%%%%%%%%%%%%%%%%%%%%%%%%%%%%%%%%%%%%%%
\begin{equation}\label{eq:discussion:ff2}
\frac{1}{c^2}\frac{\partial^2}{\partial t^2}\left(\vect{u}_\perp-\theta u_\|\delta\vect{e}\right)
-\frac{\partial^2}{\partial z^2}\left(\vect{u}_\perp-\theta u_\|\delta\vect{e}\right)
=0.
\end{equation}
%%%%%%%%%%%%%%%%%%%%%%%%%%%%%%%%%%%%%%%%%
The quantity $\vect{u}_\perp-\theta u_\|\delta\vect{e}$ that satisfies the wave equation in vacuum can be interpreted as the electric field $\vect{E}_{\perp,xy}$ associated with the daughter Alfv\'en wave perpendicular to the background magnetic field $\vect{B}_0$.
By multiplying both sides of equation (\ref{eq:discussion:ff2}) by $\beta_A\vect{B}_0$ as an outer product from the left side, we obtain
%%%%%%%%%%%%%%%%%%%%%%%%%%%%%%%%%%%%%%%%%
\begin{equation}\label{eq:discussion:ff22}
\frac{1}{c^2}\frac{\partial^2 \vect{E}_{\perp,xy}}{\partial t^2}-\frac{\partial^2 \vect{E}_{\perp,xy}}{\partial z^2}=0,
\end{equation}
%%%%%%%%%%%%%%%%%%%%%%%%%%%%%%%%%%%%%%%%%
where 
%%%%%%%%%%%%%%%%%%%%%%%%%%%%%%%%%%%%%%%%%
\begin{equation}\label{eq:discussion:ff2:sub}
\vect{E}_{\perp,xy}=
-\left(\vect{\beta_\perp}\times\vect{B_0}+\vect{\beta_\|} \times\delta\vect{B} \right).
\end{equation}
%%%%%%%%%%%%%%%%%%%%%%%%%%%%%%%%%%%%%%%%%
This corresponds to the electric field in the $xy$-plane, except for the one associated with the parent Alfv\'en wave (see equation (\ref{eq:app:Exy})).
In this way, equations (\ref{eq:norm:1st_momxy}) and (\ref{eq:norm:1st_ind}) describing daughter Alfv\'en waves reduce to the wave equation of free electromagnetic waves in the limit $\sigma\to \infty$.
On the other hand, the wave equation for acoustic waves still has a non-zero source term in the limit of $\sigma\to\infty$:
%%%%%%%%%%%%%%%%%%%%%%%%%%%%%%%%%%%%%%%%%
\begin{equation}\label{eq:discussion:ffsound1}
\frac{1}{c^2}\frac{\partial^2 \epsilon_\|}{\partial t^2}-\beta_s^2\frac{\partial^2 \epsilon_\|}{\partial z^2}
=
-w_0\left[
\frac{1}{c^2}\frac{\partial^2}{\partial t^2}\left(\delta \vect{\beta}\cdot\vect{\beta}_\perp\right)
-\frac{\partial^2}{\partial z^2}\left(\frac{\delta \vect{B}\cdot\vect{b}_\perp}{B_0^2}\right)
\right],
\end{equation}
%%%%%%%%%%%%%%%%%%%%%%%%%%%%%%%%%%%%%%%%%
%%%%%%%%%%%%%%%%%%%%%%%%%%%%%%%%%%%%%%%%%
\begin{equation}\label{eq:discussion:ffsound2}
\frac{1}{c^2}\frac{\partial^2 \beta_\|}{\partial t^2}-\beta_s^2\frac{\partial^2 \beta_\|}{\partial z^2}
=
-\frac{1}{c}\frac{\partial^2}{\partial t \partial z}
\left(
\frac{\delta \vect{B}\cdot\vect{b}_\perp}{B_0^2}
-\beta_s^2\delta \vect{\beta}\cdot\vect{\beta}_\perp
\right).
\end{equation}
%%%%%%%%%%%%%%%%%%%%%%%%%%%%%%%%%%%%%%%%%
Although the acoustic waves are generated by the interaction between parent and daughter Alfv\'en waves even for $\sigma\to\infty$, the PDI is considered to be suppressed because the acoustic wave is not possible to excite daughter Alfv\'en waves.

In this work, we have treated the amplitude $\eta$ of the parent Alfv\'en wave as constant and calculated the growth rate of infinitesimal amplitude daughter waves.
Consequently, the decay rate of the parent wave cannot be determined from this analysis.
A formulation that considers the case $\epsilon\sim\eta$ is necessary for a more precise derivation.
However, according to the general consideration of three-wave interactions, there is an implication that the decay rate of the parent wave differs only by a logarithmic factor from the growth rate of the daughter wave, namely the decay time scale of the parent wave can be written as
%%%%%%%%%%%%%%%%%%%%%%%%%%%%%%%%%%%%%%%%%
\begin{equation}
t_{\rm decay}\sim \Gamma_{\rm max}^{-1}\ln \left(\eta/\epsilon_{{\rm d},0}\right),
\end{equation}
%%%%%%%%%%%%%%%%%%%%%%%%%%%%%%%%%%%%%%%%%
where $\epsilon_{{\rm d},0}$ is the normalized initial amplitude of the daughter waves (e.g., for the Alfv\'en wave, $\epsilon_{{\rm d},0}=b_{\perp}/B_0$ at the initial state) \cite{1969npt..book.....S}.
Therefore, the growth rate obtained in this study is effective for estimating the decay rate of the parent wave.

We can apply our results to relativistically magnetized plasmas, such as the plasma of the magnetosphere of a neutron star.
Let us estimate the timescale for the decay of Alfv\'en waves generated on the surface of a neutron star, assuming they serve as the energy source for the X-ray flare of SGR1935+2154 and FRB 20200428A \cite{2020ApJ...898L..29M,2021NatAs...5..408Y,2021NatAs...5..401T,2021NatAs...5..378L,2021NatAs...5..372R,2022ApJ...931...56L,2022ApJS..260...24C}.
The estimated amplitude of the parent wave, based on the luminosity $L_{\rm X}\sim10^{41}~{\rm erg~s}^{-1}$ of the X-ray burst (i.e., $L_{\rm X}\sim4\pi R_{\rm X}^2 c \eta^2 B_{\rm NS}^2/8\pi$ where $R_{\rm X}\sim10^4~{\rm cm}$ is the size of the energy release region\footnote{
The energy release might occur in one part of the stellar surface rather than the entire surface.
Here we adopt a radius of $R_{\rm X}\sim10^4~{\rm cm}$ which is estimated by using the luminosity $L_{\rm X}\sim10^{41}~{\rm erg~s}^{-1}$ and effective temperature $T_{\rm eff}\sim80~{\rm keV}$ of the X-ray burst \cite{2020ApJ...904L..15I}.
} \cite{2020ApJ...904L..15I}, and $B_{\rm NS}\sim 10^{15}~{\rm G}$ is the surface magnetic field), is $\eta\sim3\times10^{-4}$ \cite{2020MNRAS.494.2385K,2022ApJ...933..174Y}.
Although the magnetization parameter $\sigma$ in the magnetosphere is not precisely known and depends on the efficiency of particle production in the magnetosphere, an estimation of $\sigma\sim 10^4$ has been given by previous studies \cite{1983AIPC..101..163A,2001ApJ...547..437L,2010MNRAS.406.1379M,2010MNRAS.408.2092T,2019ApJ...877...53H,2020ApJ...896..142B} \footnote{
If the $e^\pm$ fireball is formed in the X-ray burst, $\sigma$ should be much smaller \cite{2020ApJ...904L..15I}
}.
% Considering that the electrons and positrons comprising the magnetosphere have low internal energy due to radiative cooling, we evaluate the growth rate using the expression for $\theta\ll1$ (see equation (\ref{eq:growthrate_smalltheta})).
In the magnetosphere, where there are a large number of photons from the X-ray burst, we can expect the temperature of the magnetosphere to be $T\sim \mathcal{O}({\rm keV})$ due to the Compton heating \cite{2023ApJ...943L..21S}, so that we evaluate the sound speed of magnetospheric plasma as $\beta_{\rm s}\sim(1~{\rm keV}/511~{\rm keV})^{1/2}\sim0.05$.
According to equation (\ref{eq:small:theta_c}), since $\theta_c\sim8\times10^{-5}$, we find that $\theta\gg\theta_c$ and the growth rate is appropriate to be estimated by equation (\ref{eq:growthrate_gen}).
This suggests that the Alfv\'en waves propagate approximately $2\times 10^4$ wavelengths before decaying.
Assuming a wave period comparable to the rotation period of the central neutron star, $\omega_0\sim1$ Hz, $\lambda\sim10^{10}$ cm, it implies that the waves do not decay significantly within the light cylinder.
However, if we consider waves excited at a frequency similar to the observed FRB frequency (around $10^9$ Hz), the waves would decay at a distance of approximately $7\times10^5$ cm, indicating limited escape from the magnetosphere.
While these estimates are rough approximations, our results may be crucial for discussing the radiation mechanisms of FRBs.

In this study, we analyze the idealized setting of an infinitely extended uniform plane wave in a dissipationless system.
However, when applying these results to real systems, it is essential to evaluate the impacts of the following effects.
The first is the finite bandwidth effect of the parent wave.
Although stability analysis is conducted for monochromatic waves in this study, it is more natural to consider waves excited in general systems as wave packets with a finite bandwidth $\Delta \omega$.
In the context of the parametric instability, it is known that when the bandwidth of the parent wave is larger than the growth rate of daughter waves in the monochromatic case ($\Delta\omega\gg\Gamma_{\rm max}$), the effective growth rate of the daughter waves becomes $\Gamma_{\rm eff}\sim\Gamma_{\rm max}^2/\Delta \omega$, suppressed by $\Gamma_{\rm max}/\Delta \omega$ \cite{1974PhFl...17..849T}.
On the other hand, if $\Delta\omega\ll\Gamma_{\rm max}$, the phase of the parent wave and the daughter waves can be considered synchronized during the time scale of daughter wave growth, allowing us to use the analysis of the monochromatic parent wave.
The second is the effect of dissipation in the plasma.
The dominant dissipation process in the plasma can vary depending on the system under consideration.
Here, we assume an appropriate timescale for the dissipation process as $\Gamma_{\rm diss}^{-1}$.
For instability to occur in the wave, the timescale of instability without considering dissipation should be shorter than the dissipation timescale (i.e., $\Gamma_{\rm max}>\Gamma_{\rm diss}$) \cite{1974PhFl...17..849T}.
In this case, the growth rate of instability can be estimated as $\Gamma_{\rm eff}\sim \Gamma_{\rm max}-\Gamma_{\rm diss}$.
When the effect of dissipation is weak, so that the dissipation proceeds over a much longer timescale compared to the growth timescale of the instability, it is justified to use the growth rate obtained from the analysis of the monochromatic parent wave.

In this study, we have only investigated the PDI within the linear stage (i.e., $\mathcal{O}(\varepsilon)\ll1$), so that further research is necessary to understand the development in the nonlinear stage.
In systems where $\sigma \gg 1$, meaning the electromagnetic energy density greatly exceeds the matter energy density, non-thermal high-energy phenomena such as non-thermal particle acceleration may occur due to the transfer of electromagnetic energy to low-density regions.
Additionally, for the generated acoustic waves, damping effects including the Landau damping are expected to occur, thus requiring an analysis incorporating kinetic effects to further quantitatively understand the PDI.
Magnetohydrodynamic simulation encounters difficulty in dealing with large $\sigma$ and cannot analyze kinetic phenomena such as particle acceleration, making Particle-in-Cell (PIC) simulations more suitable for such studies.
While this study also applied linearization to $\eta$, further extensions are possible.
Equations (19)-(22) are formulated for general $\eta$, and we are currently conducting ongoing research to directly analyze these equations.
When $\eta$ reaches unity, higher-order wave-wave interactions (e.g., 4-wave interaction) may become significant, although this phenomenon lies beyond the scope of this study.
Exploring higher-order wave-wave interactions through simulations, particularly PIC simulations, will provide valuable insights into understanding the stability of Alfv\'en waves.

\section{Summary}
We have investigated the parametric decay instability (PDI) of circularly polarized Alfv\'en waves in a plasma with relativistic magnetization, i.e., $\sigma\gg1$, based on the special relativistic magnetohydrodynamic (MHD) equations.
Our results indicate that even in plasmas with relativistic magnetization, the Alfv\'en wave is unstable to three-wave interactions where the parent wave generates two daughter waves - a forward propagating acoustic wave and a backward propagating Alfv\'en wave.
We have obtained the analytical expression of the maximum growth rate (see equation (\ref{eq:growthrate_sig})) as follows:
%%%%%%%%%%%%%%%%%%%%%%%%%%%%%%%%%%%%%%%%%
\begin{equation*}
\Gamma_{\rm max}/\omega_0=\frac{1}{2}\eta\theta^{-1/2}\frac{\sqrt{1-\theta}}{1+\theta}\frac{1+\sigma\left(1+\theta^2\right)}{\left(1+\sigma\right)^{3/2}},
\end{equation*}
%%%%%%%%%%%%%%%%%%%%%%%%%%%%%%%%%%%%%%%%%
where $\eta=\delta B/B_0$ is the amplitude of the parent Alfv\'en wave, $\sigma=B_0^2/(4\pi(\epsilon_0+p_0))$ is the magnetization, $\theta=\beta_{\rm A}/\beta_{\rm s}$ is the ratio of the sound speed to the Alfv\'en velocity, and $\omega_0$ is the frequency of the parent Alfv\'en wave.
This is only valid when the ratio $\theta$ of the sound speed to Alfv\'en velocity is greater than $\theta_c$, which is expressed by equation (\ref{eq:small:theta_c}).
For $\theta\ll\theta_c$, the growth rate can be written as follows:
%%%%%%%%%%%%%%%%%%%%%%%%%%%%%%%%%%%%%%%%%
\begin{equation*}
\Gamma_{\rm max}/\omega_0=\frac{\sqrt{3}}{2} \eta^{2 / 3} \left(1+\sigma\right)^{-1 / 3}.
\end{equation*}
%%%%%%%%%%%%%%%%%%%%%%%%%%%%%%%%%%%%%%%%%
As can be seen from these equations, we have observed that the instability is suppressed with increasing magnetization $\sigma$.
Our results for the instability growth rates are consistent with the known non-relativistic limits.
Such a process may provide a clue to the energy conversion mechanism from electromagnetic field to plasma in environments around compact objects, where strongly magnetized plasmas are expected to exist, e.g., in the magnetospheres of neutron stars or black holes.

\begin{acknowledgments}
We are grateful to 
Toshio Terasawa, Kenji Toma, Takayoshi Sano, Yohei Kawazura, Shuichi Matsukiyo, Masanori Iwamoto, Shoma Kamijima, Rei Nshiura, Koutarou Kyutoku, Tomoki Wada, Shota Kisaka, Kazumi Kashiyama, and Shuta Tanaka
for fruitful discussion and valuable comments.
This work is supported by Grants-in-Aid for Scientific
Research No. 23H01172, 23H05430, 23H04900, 22H00130, 20H01901, 20H01904, 20H00158 (KI), 21J01450 (WI) from the Ministry
of Education, Culture, Sports, Science and Technology (MEXT) of Japan.

\end{acknowledgments}

\appendix

\section{Derivation of linearlized equations}\label{app:derivation}

\subsection{Fluid equations with electromagnetic field as external field}
In order to derive equations (\ref{eq:full:1})-(\ref{eq:full:4}), we first simplify equations (\ref{eq:basic:energy})-(\ref{eq:basic:ind_eq}).
Using the Maxwell equations, the energy-momentum conservation law can be rewritten into the fluid equations in the external electromagnetic field as follows:
%%%%%%%%%%%%%%%%%%%%%%%%%%%%%%%%%%%%%%%%%%%%%%%
\begin{equation}\label{eq:app:ene_ext}
	\frac{1}{c}\frac{\partial}{\partial t}\left[(\epsilon+p) \gamma^{2}-p\right]
	+\pardel{}{z}\left[(\epsilon+p) \gamma^{2} \beta_z\right]=\frac{1}{c}\vect{j}\cdot\vect{E},
\end{equation}
%%%%%%%%%%%%%%%%%%%%%%%%%%%%%%%%%%%%%%%%%%%%%%%
%%%%%%%%%%%%%%%%%%%%%%%%%%%%%%%%%%%%%%%%%%%%%%%
\begin{equation}\label{eq:app:mom_ext}
	\frac{1}{c}\frac{\partial}{\partial t}[(\epsilon+p)\left.\gamma^{2} \boldsymbol{\beta}\right]
	+\pardel{}{z}\left[(\epsilon+p) \gamma^{2} \beta_z \boldsymbol{\beta}\right]
    +\pardel{p}{z}
    =\rho_e\vect{E}+\frac{1}{c}\vect{j}\times\vect{B},
\end{equation}
%%%%%%%%%%%%%%%%%%%%%%%%%%%%%%%%%%%%%%%%%%%%%%%
where $\rho_e$ is the charge density and $\vect{j}$ is the current density.
The electromagnetic field is calculated from the Maxwell equations and the ideal MHD conditions.
Since the electric field satisfies the ideal MHD condition $\vect{E}=-\vect{\beta}\times\vect{B}$, the evolution of the magnetic field from the Maxwell-Faraday equation is described by the following induction equation:
%%%%%%%%%%%%%%%%%%%%%%%%%%%%%%%%%%%%%%%%%%%%%%%
\begin{equation}\label{eq:app:ind_eq}
	\frac{1}{c}\pardel{\vect{B}}{t}=\nabla\times\left(\vect{\beta}\times\vect{B}\right).
\end{equation}
%%%%%%%%%%%%%%%%%%%%%%%%%%%%%%%%%%%%%%%%%%%%%%%
The charge density $\rho_e$ and the current density $\vect{j}$ in the right-hand side of equations (\ref{eq:app:ene_ext}) and (\ref{eq:app:mom_ext}), respectively, can be written as follows by using Maxwell's equations:
%%%%%%%%%%%%%%%%%%%%%%%%%%%%%%%%%%%%%%%%%%%%%%%
\begin{equation}\label{eq:app:rhoe}
	\rho_e
	=\frac{1}{4\pi}\nabla\cdot\vect{E},
	% =-\frac{1}{4\pi}\nabla\cdot\left(\vect{\beta}\times\vect{B}\right)
\end{equation}
%%%%%%%%%%%%%%%%%%%%%%%%%%%%%%%%%%%%%%%%%%%%%%%
%%%%%%%%%%%%%%%%%%%%%%%%%%%%%%%%%%%%%%%%%%%%%%%
\begin{equation}\label{eq:app:j}
	\vect{j}
    % =\frac{c}{4\pi}\left[\frac{1}{c}\pardel{}{t}\left(\vect{\beta}\times\vect{B}\right)+\nabla\times \vect{B}\right]
    =\frac{c}{4\pi}\left(\nabla\times \vect{B}-\frac{1}{c}\pardel{\vect{E}}{t}\right).
\end{equation}
%%%%%%%%%%%%%%%%%%%%%%%%%%%%%%%%%%%%%%%%%%%%%%%
Note that since the electric field $\vect{E}$ satisfies the ideal MHD conditions, the charge density $\rho_e$ and the current density $\vect{j}$ can be obtained from the magnetic field $\vect{B}$ and velocity $\vect{\beta}$.

\subsection{Derivation of linearized equations}
The quantities required in the subsequent calculations up to the order of $\mathcal{O}(\varepsilon)$ are calculated as follows.
Square of the Lorentz factor of the fluid:
%%%%%%%%%%%%%%%%%%%%%%%%%%%%%%%%%%%%%%%%%%%%%%%
\begin{equation}\label{eq:app:gam2}
	\gamma^2=\frac{1}{1-\vect{\beta}\cdot\vect{\beta}}
	% =\left[1-\left(\delta \vect{\beta}+\vect{\beta_\perp}+\vect{\beta_\|}\right)^2\right]^{-1}
	% =\delta\gamma^2\left(1-2\delta\gamma^2\delta \vect{\beta}\cdot\vect{\beta_\perp}\right)^{-1}
	\sim\delta\gamma^2\left(1+2\delta\gamma^2\delta \vect{\beta}\cdot\vect{\beta_\perp}\right),
\end{equation}
%%%%%%%%%%%%%%%%%%%%%%%%%%%%%%%%%%%%%%%%%%%%%%%
where $\delta\gamma=(1-\delta \beta^2)^{-1/2}=(1-\eta^2 \beta_{A,\eta}^2)^{-1/2}$.
Enthalpy density:
%%%%%%%%%%%%%%%%%%%%%%%%%%%%%%%%%%%%%%%%%%%%%%%
\begin{equation}%\label{key}
	\epsilon+p\sim\left(\epsilon_0+p_0\right)+\epsilon_\|\left(1+\beta_s^2\right).
\end{equation}
%%%%%%%%%%%%%%%%%%%%%%%%%%%%%%%%%%%%%%%%%%%%%%%
$z$ component of the electric field:
%%%%%%%%%%%%%%%%%%%%%%%%%%%%%%%%%%%%%%%%%%%%%%%
\begin{equation}\label{eq:app:Ez}
	E_z\vect{e}_z\sim-\left(\delta\vect{\beta}\times\vect{b_\perp}+ \vect{\beta_\perp}\times\delta \vect{B}\right).
\end{equation}
%%%%%%%%%%%%%%%%%%%%%%%%%%%%%%%%%%%%%%%%%%%%%%%
Note that $E_z$ is a quantity of $\mathcal{O}(\varepsilon)$.
In-plane $xy$ component of the electric field:
%%%%%%%%%%%%%%%%%%%%%%%%%%%%%%%%%%%%%%%%%%%%%%%
\begin{equation}\label{eq:app:Exy}
	\vect{E_{xy}}
	\sim-\left(\delta\vect{\beta}\times\vect{B_0}+\vect{\beta_\perp}\times\vect{B_0}+\vect{\beta_\|} \times\delta\vect{B} \right).
\end{equation}
%%%%%%%%%%%%%%%%%%%%%%%%%%%%%%%%%%%%%%%%%%%%%%%
Square of the absolute value of the electric field:
%%%%%%%%%%%%%%%%%%%%%%%%%%%%%%%%%%%%%%%%%%%%%%%
\begin{equation}\label{eq:app:E2}
	E^2\sim
	\delta\beta^2 B_0^2
	+2\left(\delta \vect{\beta}\cdot\vect{\beta_\perp}\right)B_0^2
	+2\chi_0\delta\beta\delta BB_0\beta_\|.
\end{equation}
%%%%%%%%%%%%%%%%%%%%%%%%%%%%%%%%%%%%%%%%%%%%%%%
Square of the absolute value of the magnetic field:
%%%%%%%%%%%%%%%%%%%%%%%%%%%%%%%%%%%%%%%%%%%%%%%
\begin{equation}\label{eq:app:B2}
	B^2
 % =\left(\vect{B_0}+\delta\vect{B}+\vect{b_\perp}\right)\cdot\left(\vect{B_0}+\delta\vect{B}+\vect{b_\perp}\right)
	\sim B_0^2+\delta B^2+2\delta \vect{B}\cdot\vect{b_\perp}.
\end{equation}
%%%%%%%%%%%%%%%%%%%%%%%%%%%%%%%%%%%%%%%%%%%%%%%

Using equations (\ref{eq:app:Ez})--(\ref{eq:app:B2}), calculate the right-hand side of equations (\ref{eq:app:ene_ext}) and (\ref{eq:app:mom_ext}).
First, the work rate of the electromagnetic field $\vect{j}\cdot\vect{E}$ is as follows:
%%%%%%%%%%%%%%%%%%%%%%%%%%%%%%%%%%%%%%%%%%%%%%%
\begin{equation}%\label{key}
	\vect{j}\cdot\vect{E}\sim
	\frac{c}{4\pi}
	\left[
	B_0\vect{\beta_\perp}\cdot\pardel{}{z}\left(\delta\vect{B}\right)
	+B_0\delta\vect{\beta}\cdot\pardel{\vect{b_\perp}}{z}
	-B_0^2\frac{1}{c}\pardel{}{t}\left(\delta \vect{\beta}\cdot\vect{\beta_\perp}\right)
	-\chi_0\delta\beta\delta BB_0\frac{1}{c}\pardel{\beta_\|}{t}.
	\right]
\end{equation}
%%%%%%%%%%%%%%%%%%%%%%%%%%%%%%%%%%%%%%%%%%%%%%%
Here we used the following relations, which hold for circularly polarised Alfv\'en waves:
%%%%%%%%%%%%%%%%%%%%%%%%%%%%%%%%%%%%%%%%%%%%%%%
\begin{equation}%\label{key}
	\delta\vect{B}\cdot\pardel{}{z}\left(\delta\vect{B}\right)=0,~~
	\delta\vect{\beta}\cdot\pardel{}{z}\left(\delta\vect{B}\right)=0.
\end{equation}
%%%%%%%%%%%%%%%%%%%%%%%%%%%%%%%%%%%%%%%%%%%%%%%

Next, let us calculate the electric force $\rho_e \vect{E}$.
From the equation (\ref{eq:app:rhoe}), the charge density$\rho_e$ is as follows.
%%%%%%%%%%%%%%%%%%%%%%%%%%%%%%%%%%%%%%%%%%%%%%%
\begin{equation}%\label{key}
	\rho_e
    % =\frac{1}{4\pi}\nabla\cdot\vect{E}
	=\frac{1}{4\pi}\pardel{E_z}{z}\vect{e}_z
	\sim
	-\frac{1}{4\pi}
	\pardel{}{z}\left[
	\frac{\vect{B_0}}{B_0}\cdot\left(\delta\vect{\beta}\times\vect{b_\perp}+ \vect{\beta_\perp}\times\delta \vect{B}\right)
	\right].
\end{equation}
%%%%%%%%%%%%%%%%%%%%%%%%%%%%%%%%%%%%%%%%%%%%%%%
As can be seen from this equation, $\rho_e=\mathcal{O}(\varepsilon)$.
Since $E_z$ is also a quantity of $\mathcal{O}(\varepsilon)$, the electric force in the $z$ direction $\rho_e E_z$ is a second order of $\varepsilon$:
%%%%%%%%%%%%%%%%%%%%%%%%%%%%%%%%%%%%%%%%%%%%%%%
\begin{equation}%\label{key}
	\rho_e E_z\sim 0.
\end{equation}
%%%%%%%%%%%%%%%%%%%%%%%%%%%%%%%%%%%%%%%%%%%%%%%
The $xy$ in-plane component of the electric force is as follows:
%%%%%%%%%%%%%%%%%%%%%%%%%%%%%%%%%%%%%%%%%%%%%%%
\begin{equation}%\label{key}
	\rho_e \vect{E_{xy}}
	\sim
	\frac{1}{4\pi}
	\left(\delta\vect{\beta}\times\vect{B_0}\right)
	\pardel{}{z}\left[
		\frac{\vect{B_0}}{B_0}\cdot\left(\delta\vect{\beta}\times\vect{b_\perp}+ \vect{\beta_\perp}\times\delta \vect{B}\right)
	\right].
\end{equation}
%%%%%%%%%%%%%%%%%%%%%%%%%%%%%%%%%%%%%%%%%%%%%%%

From equation (\ref{eq:app:j}), the $z$ component of the magnetic force $\vect{j}\times \vect{B}$ can be calculated as follows:
%%%%%%%%%%%%%%%%%%%%%%%%%%%%%%%%%%%%%%%%%%%%%%%
\begin{equation}%\label{key}
\left[\frac{1}{c}\vect{j}\times\vect{B}\right]_z=
-\frac{\partial}{\partial z}\left(\frac{\delta \vect{B} \cdot \vect{b_\perp}}{4\pi}\right)
+\frac{B_{0}}{4\pi}\left[
\vect{\delta B} \cdot  \frac{1}{c}\frac{\partial \vect{\beta_{\perp}}}{\partial t}
+\vect{b_{\perp}} \cdot \frac{1}{c}\frac{\partial}{\partial t}(\delta \vect{\beta})
\right]
-\frac{\delta B^{2}}{4\pi}  \frac{1}{c}\frac{\partial {\beta}_{\|}}{\partial t}.
\end{equation}
%%%%%%%%%%%%%%%%%%%%%%%%%%%%%%%%%%%%%%%%%%%%%%%
The $xy$ in-plane component of the magnetic force is as follows:
%%%%%%%%%%%%%%%%%%%%%%%%%%%%%%%%%%%%%%%%%%%%%%%
\begin{multline}\label{eq:app:jxBxy}
\left[\frac{1}{c}\vect{j}\times\vect{B}\right]_{xy}=
\frac{B_{0}}{4\pi}\frac{\partial}{\partial z}\left(\delta \vect{B}+\vect{b_\perp}\right)
-\frac{B_{0}^{2}}{4\pi} \frac{1}{c} \frac{\partial}{\partial t}(\delta \vect{\beta})
-\frac{B_{0}^{2}}{4\pi} \frac{1}{c} \frac{\partial \vect{\beta_{\perp}}}{\partial t}
+\frac{B_0}{4\pi} \frac{1}{c} \frac{\partial}{\partial t}\left(\beta_{\|} \delta \vect{B}\right)
-\frac{\delta \vect{B} \cdot \vect{b_{\perp}}}{4\pi}\frac{1}{c}\frac{\partial}{\partial t}(\delta \vect{\beta})\\
-\delta \vect{\beta}\left(\frac{\delta \vect{B}}{4\pi} \cdot \frac{1}{c}\frac{\partial \vect{b_{\perp}}}{\partial t}\right)
-\chi_0\frac{\delta B}{4\pi}\delta \beta \frac{1}{c}\frac{\partial \vect{b_{\perp}}}{\partial t}
-\frac{\delta B^{2}}{4\pi} \frac{1}{c}\frac{\partial \vect{\beta_\perp}}{\partial t}
+\frac{\delta \vect{B}}{4\pi}\left(\frac{1}{c}\frac{\partial \vect{\beta_{\perp}}}{\partial t} \cdot \delta \vect{B}\right)
+\frac{\delta \vect{B} \cdot \vect{\beta_{\perp}}}{4\pi} \frac{1}{c}\frac{\partial}{\partial t}(\delta \vect{B}).
\end{multline}
%%%%%%%%%%%%%%%%%%%%%%%%%%%%%%%%%%%%%%%%%%%%%%%
By substituting the above equations (\ref{eq:app:gam2})--(\ref{eq:app:jxBxy}) into equations (\ref{eq:app:ene_ext}), (\ref{eq:app:mom_ext}) and (\ref{eq:app:ind_eq}), equations (\ref{eq:full:1})--(\ref{eq:full:4}) can be obtained by straightforward calculation.

% \nocite{*}

\bibliography{apssamp}% Produces the bibliography via BibTeX.

\end{document}